\documentclass{amsproc}
\usepackage{amssymb,xspace,a4wide}

\DeclareFontFamily{U}{rsf}{}
\DeclareFontShape{U}{rsf}{m}{n}{
  <5> <6> rsfs5 <7> <8> <9> rsfs7 <10->  rsfs10}{}
\DeclareMathAlphabet{\mathscr}{U}{rsf}{m}{n}

\theoremstyle{definition}

\theoremstyle{remark}

\numberwithin{equation}{section}
\newlength{\picwidth} 
\newlength{\miniwidth} 
\newlength{\nanowidth} 
\newlength{\melowidth} 
\newlength{\leftminiwidth} 
\newlength{\rightminiwidth} 
\newlength{\minipagewidth} 
\numberwithin{equation}{section}

\def\varpi{t}

\def\Im{\,{\rm Im}\,}

\def\({\left(}
\def\){\right)}
\def\[{\left[}
\def\]{\right]}
\def\<{\left\langle}
\def\>{\right\rangle}
\def\hf{{1\over 2}}

\renewcommand{\d}{\mathrm{d}}
\newcommand{\de}{\mathrm{d}}

\newcommand{\I}{\mathrm{i}}

\newcommand{\cL}{\mathcal{L}}
\newcommand{\cD}{\mathcal{D}}

\newcommand{\p}{\partial}

\newcommand{\half}{\frac{1}{2}}

\newcommand{\cV}{\mathcal{V}}

\newcommand{\cS}{\mathcal{S}}

\newcommand{\cK}{\mathcal{K}}
\newcommand{\cM}{\mathcal{M}}
\newcommand{\cW}{\mathcal{W}}
\newcommand{\cN}{\mathcal{N}}

\newcommand{\cX}{\mathcal{X}}

\newcommand{\cT}{\mathcal{T}}
\newcommand{\cJ}{\mathcal{J}}

\newcommand{\cB}{\mathcal{B}}

\DeclareSymbolFont{AMSa}{U}{msa}{m}{n}
\DeclareSymbolFont{AMSb}{U}{msb}{m}{n}
\DeclareMathSymbol{\fieldR}{\mathalpha}{AMSb}{"52}

\newcommand{\N}{{\mathcal N}}
\newcommand{\kahler}{{K\"ahler}\xspace}
\newcommand{\hk}{{hyperk\"ahler}\xspace}
\newcommand{\qk}{{quaternion-K\"ahler}\xspace}

\newcommand{\cZ}{\mathcal{Z}}

\newcommand{\cO}{\mathcal{O}}
\newcommand{\cH}{\mathcal{H}}
\newcommand{\cU}{\mathcal{U}}
\newcommand{\cA}{\mathcal{A}}

\newcommand{\pa}{\partial}
\newcommand{\nn}{\nonumber}

\newcommand{\eps}{\epsilon}

\newcommand{\IR}{\mathbb{R}}
\newcommand{\IC}{\mathbb{C}}
\newcommand{\IZ}{\mathbb{Z}}

\newcommand{\sgn}{\mbox{sgn}}
\newcommand{\tzeta}{\tilde\zeta}

\newcommand{\txi}{\tilde\xi}

\newcommand{\CP}{\mathbb{P}^1}

\def\bea{\begin{eqnarray}}
\def\eea{\end{eqnarray}}
\def\be{\begin{equation}}
\def\ee{\end{equation}}
\def\ba{\begin{align}}
\def\ea{\end{align}}
\def\bse{\begin{subequations}}
\def\ese{\end{subequations}}

\def\ba{\bar a}

\def\bz{\bar z}

\def\bZ{\bar Z}

\newcommand{\Li}{{\rm Li}}

\def\XXint#1#2#3{{\setbox0=\hbox{$#1{#2#3}{\int}$}
\vcenter{\hbox{$#2#3$}}\kern-.5\wd0}}

\def\cij#1{c}
\def\ci#1{c}

\def\ellg#1{\ell_{#1}}

\def\Fcl{F^{\rm cl}}

\def\bH{\overline{H}}
\def\bOm{\overline{\Omega}}

\newcommand{\expe}[1]{{\bf E}\!\left( #1\right)}

\DeclareMathOperator{\Td}{Td}
\DeclareMathOperator{\ch}{ch}

\newcommand{\beq}{\begin{eqnarray}}
\newcommand{\eeq}{\end{eqnarray}}

\def\cla{\tilde c_a}
\def\cl0{\tilde c_0}

\newcommand{\bfb}{{\boldsymbol b}}
\newcommand{\bfc}{{\boldsymbol c}}

\newcommand{\bfk}{{\boldsymbol k}}
\newcommand{\bfp}{{\boldsymbol p}}
\newcommand{\bfq}{{\boldsymbol q}}

\newcommand{\bft}{{\boldsymbol t}}
\newcommand{\bfz}{{\boldsymbol z}}

\newcommand{\bfy}{{\boldsymbol y}}
\newcommand{\bfeps}{{\boldsymbol \epsilon}}
\newcommand{\bfmu}{{\boldsymbol \mu}}

\newcommand{\bfch}{{\bf{\boldsymbol c}_2}}
\newcommand{\bfcl}{\tilde{\boldsymbol c}}

\newcommand{\bftxip}{\smash{\tilde{\boldsymbol\xi}}\vphantom{\txi}'}
\newcommand{\bfxi}{{\boldsymbol \xi}}

\newcommand{\gammap}{\gamma}

\def\trans{g}
\def\hgamma{\hat\gamma}

\def\ZTheta{\Xi}
\def\BTheta{\Upsilon}

\def\kbp{(k+b)_+}

\def\signkp{(-1)^{\bfk\cdot\bfp}}

\def\qrD{\lambda_{\rm D}(\gamma)}
\def\qrDp{\lambda_{\rm D}(\gamma')}
\def\qrDg#1{\lambda_{\rm D}(\gamma_{#1})}

\def\todaQ{T}

\begin{document}

\title[Quantum hypermultiplet moduli spaces in $\cN=2$ string vacua: a review]{
Quantum hypermultiplet moduli spaces \\ in $\cN=2$ string vacua: a review}

\author[]{Sergei Alexandrov}
\address{Universi\'e Montpellier 2, Laboratoire Charles Coulomb, F-34095,
Montpellier, France}
\email{salexand@univ-montp2.fr}

\author[]{Jan Manschot}
\address{Bethe Center for Theoretical Physics, Bonn University, Nu\ss allee 12, 53115 Bonn, Germany}
\address{Max Planck Institute for Mathematics, Vivatsgasse 7, 53111 Bonn, Germany}
\email{manschot@uni-bonn.de}

\author[]{Daniel Persson}
\address{Fundamental Physics, Chalmers University of Technology,
  412 96, Gothenburg, Sweden}
  \email{daniel.persson@chalmers.se}

\author[]{Boris Pioline}

\address{CERN PH-TH, Case C01600, CERN, CH-1211 Geneva 23, Switzerland}

\address{Laboratoire de Physique Th\'eorique et Hautes Energies, CNRS UMR 7589,
         and Universit\'e Pierre et Marie Curie - Paris 6, 4 place Jussieu,
         75252 Paris cedex 05, France
}

\email{boris.pioline@cern.ch, pioline@lpthe.jussieu.fr}

  \thanks{{\it Prepared for the proceedings of  String Math 2012, Bonn.
  }}

\begin{abstract}
The hypermultiplet moduli space $\cM_H$ in type II string theories compactified on a Calabi-Yau threefold $X$
is largely constrained by supersymmetry (which demands
quaternion-K\"ahlerity),  S-duality (which requires an isometric action of $SL(2,\IZ)$) and
regularity. Mathematically,  $\cM_H$ ought to encode all
generalized Donaldson-Thomas invariants on $X$
consistently with wall-crossing, modularity and homological mirror symmetry.
We review recent progress towards computing the exact metric on $\cM_H$,
or rather the exact complex contact structure on its twistor space.
\end{abstract}

\maketitle

\tableofcontents
%\bigskip
\vspace{-.5cm}
\begin{flushright}
L2C:13-038,
Bonn-TH-2013-05,
CERN-PH-TH/2013-048\\
arXiv:1304.0766v3, revised Jan 2015.
\end{flushright}

\newpage
%===========================================================

\section{Introduction}

String vacua with $N=2$ supersymmetry in four dimensions offer a unique opportunity to
investigate non-perturbative aspects of the low energy effective action and of the spectrum
of black hole bound states. Unlike in vacua with higher supersymmetry, the two-derivative
effective action  in general receives non-trivial quantum corrections, while degeneracies
of BPS black holes  depend non trivially on the value of the moduli at spatial infinity. Both issues are
in fact related, since BPS black holes in 4 dimensions yield BPS instantons upon reduction on a circle,
and the resulting instanton corrections to the three-dimensional effective action can sometimes
(after T-duality along the circle) lift back to 4 dimensions.

For ungauged $\cN=2$ vacua, the complete two-derivative effective action is encoded in
the  Riemannian metric on the moduli space, which famously factorizes as the product
$\cM_V \times \cM_H$ of the vector multiplet (VM) and hypermultiplet (HM) moduli spaces,
respectively \cite{Bagger:1983tt,deWit:1984px}. A complete understanding of the former
was achieved in the 90s, leading to deep connections with algebraic geometry, most notably
the discovery of classical mirror symmetry. By contrast,  our understanding of the latter
has long remained rudimentary, mainly due to the difficulty of parametrizing  \qk (QK)
metrics on $\cM_H$, as required by supersymmetry. The situation has considerably
improved in recent years, as twistorial techniques
\cite{MR664330,Karlhede:1984vr,Hitchin:1986ea,MR1001707, MR1096180,deWit:2001dj,Lindstrom:2008gs} were used to
reformulate this problem analytically, in terms of the complex contact structure on the twistor
space $\cZ$ (a $\CP$-bundle over $\cM_H$), and a suitable set of complex Darboux coordinates on
$\cZ$.

The purpose of this contribution is to give a survey
of recent progress towards determining the exact  hypermultiplet moduli space metric
(see \cite{Alexandrov:2011va} for a review with different emphasis).
We  focus on type II strings compactified on Calabi-Yau (CY) threefolds ,
although other dual formulations of the same vacua (see \cite{Louis:2011aa,
Alexandrov:2012pr,Alexandrov:2014jua}
for recent progress on $K3\times T^2$ heterotic vacua) may eventually be useful for achieving the stated goal.
In \S\ref{sec_pert}, we summarize the structure of the
perturbative hypermultiplet moduli space in type IIA and type IIB  vacua, emphasizing
its twistorial description. In \S\ref{sec_Dinst}, we discuss instanton corrections from Euclidean
D-branes wrapped on supersymmetric cycles inside the CY threefold $X$,
and provide a  twistorial construction of these corrections
parametrized by the Donaldson-Thomas invariants of $X$.
We explain the consistency of this construction with wall-crossing using the so called QK/HK correspondence
which allows to reformulate the resulting corrections to the QK metric on $\cM_H$
in terms of corrections to an auxiliary, or ``dual'', \hk (HK) space $\cM'_H$.
In \S\ref{sec_D3}, we discuss the implications of the modular symmetry of type IIB
strings (known as S-duality) for these D-instanton corrections, with particular emphasis on
D3-brane instantons (corresponding to divisors in $X$). We show that in the large volume, one-instanton
approximation, S-duality holds thanks to special modular properties of the DT invariants
for divisors, and of the indefinite theta series which sum up D1-D(-1) instanton effects at fixed D3-brane charge.
In \S\ref{sec_NS5}, we use the same duality to obtain Neveu-Schwarz (NS) fivebrane  instantons
from D5-instantons, and relate these contributions to the topological string
amplitude on $X$. We also discuss some conjectural relations between NS5-branes and quantum integrable systems.

\noindent{\it Acknowledgments.} We are grateful to P. Roche, F. Saueressig and S. Vandoren
for collaboration on some of the material presented here, and to N.~Hitchin, A.~Kleinschmidt,
S.~Monnier, R.~Minasian, G.~Moore, A.~Neitzke, B.~Nilsson, and
Y.~Soibelman for related discussions. DP and BP wish to thank the
organizers of String Math 2012 for the opportunity to report on some
of this work. DP also thanks the organizers and participants of the mini-workshop on ``Hypers'' in Hamburg,
March 2013, where a series of lectures on part of this
work was given. JM is supported in part by a Krupp fellowship.

\medskip

\noindent{\it Note added in proof, Jan 2015.} This review was updated to incorporate references to relevant 
works which have appeared since the first release in April 2013.

\section{Perturbative moduli space\label{sec_pert}}

In this section we discuss the one-loop  corrected
hypermultiplet moduli space in type IIA and type IIB string theories compactified on CY threefolds $X$ and $\hat X$,
respectively. Higher loop corrections are expected to vanish after suitable field redefinitions
\cite{Gunther:1998sc,Robles-Llana:2006ez,Alexandrov:2008gh}.
If $(X,\hat X)$ is a mirror pair, then the  moduli spaces are isometric as a consequence of
classical mirror symmetry \cite{Candelas:1990rm}.

\subsection{Type IIA}

\subsubsection{Topology}
\label{sec_topology}

Type IIA string theory associates to each compact
CY threefold $X$ a real $4(h_{2,1}(X)+1)$-dimensional quaternion-K\"ahler space
$\cM_H=\cM_H(X)$. Topologically, $\cM_H$ is a $\mathbb{C}^{\times}$-bundle
\be
\mathbb{C}^{\times}\, \longrightarrow\,  \cM_H(X)\,  \longrightarrow\,   \cJ_W(X)
\,  \longrightarrow\,  \cM_C(X)
\label{IIAbundle}
\ee
over the Weil intermediate Jacobian $\cJ_W(X)$. The latter
is a torus bundle over the complex structure moduli space $\cM_C(X)$,
with  generic fiber  $\cT=H^3(X, \mathbb{R})/H^3(X, \mathbb{Z})$,
endowed with the Weil complex structure where $H^{3,0}\oplus H^{1,2}$ generate the
holomorphic tangent space.

To see how this arises from physics, consider the
$\mathbb{C}^{\times}$-bundle $\cL_X\to \cM_C(X)$ with fibre the space
of nowhere vanishing holomorphic 3-forms $\Omega^{3,0}$ on $X$.
Fixing  a symplectic basis
$(A^{\Lambda}, B_{\Lambda})$, $\Lambda=0,\dots, h_{2,1}(X)$ of $\Gamma=H_3(X, \mathbb{Z})$, the period integrals
\be
X^{\Lambda}=\int_{A^{\Lambda}} \Omega^{3,0},
\qquad
F_{\Lambda} = \int_{B_{\Lambda}} \Omega^{3,0}
\ee
realize $\cL_X$ as a complex Lagrangian cone in $H^3(X, \mathbb{C})$. Locally, the
$B$-periods $F_\Lambda$ can be expressed in terms of the $A$-periods $X^\Lambda$
as derivatives of a holomorphic prepotential $F(X^\Lambda)$ homogeneous of degree 2.
The ratios $z^a=X^{a}/X^0, \, a=1, \dots, h_{2,1}$, parametrize the moduli space of complex
structures $\cM_C(X)$, and describe the scalar degrees of freedom in type IIA$/X$ originating
from the metric in 10 dimensions. The periods $(X^\Lambda,F_\Lambda)$ are valued in the
Hodge bundle $\cL_X$ (times a symplectic vector bundle associated to changes
of the symplectic basis).

In addition, the periods of the ten-dimensional Ramond-Ramond (RR) three-form $C$
\be
\label{defC}
\zeta^{\Lambda} = \int_{A^\Lambda} C, \qquad \qquad \tilde\zeta_\Lambda = \int_{B_{\Lambda}} C
\ee
yield  scalar moduli valued in $H^3(X, \mathbb{R})$. Invariance under large gauge transformations $C\to C+H$ with $H\in \Gamma$
imply that $(\zeta^{\Lambda}, \tilde\zeta_\Lambda)$ are periodic with integer periods, hence
live in  the torus $\cT$.
Sometimes we abuse notation and write $\Omega^{3,0}=(X^\Lambda,F_\Lambda)$, $C=(\zeta^{\Lambda},\tilde\zeta_\Lambda)$.
Just like $\Omega^{3,0}$, the vector $C$ transforms by a symplectic rotation under monodromies  in $\cM_C(X)$,
implying that $\cT$ is non-trivially fibered over $\cM_X$. The total space
of this bundle is the intermediate Jacobian $\cJ_W(X)$.

Finally, the four-dimensional dilaton $e^{\phi}$ and  the Poincar\'e dual $\sigma$
to the $B$-field in four dimensions provide an additional complex
scalar degree of freedom in four dimensions, corresponding to the $\IC^\times$ fiber in
\eqref{IIAbundle}.  Large gauge transformations of the $B$-field identify $\sigma\mapsto
\sigma+2\kappa$ with $\kappa$ integer  (for a suitable normalization),
while the afore-mentioned large gauge transformations  also act on the axion
$\sigma$ by a shift \cite{Alexandrov:2010np,Alexandrov:2010ca},
\be
(C, \sigma) \longmapsto \Big(C+H, \sigma + 2\kappa + \left<C, H\right> +2c(H)\Big) .
\label{heis0}
\ee
Here $c(H)$ provides the quadratic refinement $\lambda(H)\equiv (-1)^{2c(H)}$ of the intersection form
$\left<\, , \, \right>$ on $\Gamma$ satisfying
\be
\label{shiftcoc}
\lambda(H+H')=(-1)^{\left<H, H'\right>} \lambda(H)\, \lambda(H') .
\ee
Given a choice of symplectic basis of $H^{3}(X, \mathbb{Z})$, any quadratic refinement can be
parametrized as
\be
\label{lamchar}
\lambda(H)=(-1)^{2c(H)} = e^{-\I\pi m_\Lambda n^{\Lambda}+2\pi \I (m_\Lambda \theta^{\Lambda}- n^{\Lambda}\phi_\Lambda)},
\ee
where $H=(n^\Lambda,m_\Lambda)$ are the components of $H$ along $(A^\Lambda,B_\Lambda)$ and
$\Theta\equiv(\theta^{\Lambda}, \phi_\Lambda)$ are a choice of characteristics in $\cT$.
Note that \eqref{lamchar} defines $c(H)$ only modulo integers,
but the corresponding ambiguity in \eqref{heis0} can be absorbed in $\kappa$.
The extra shift $2c(H)$ of $\sigma$ in \eqref{heis0} is needed to ensure the closure of the group action.
Altogether, the large gauge transformations of the $B$ and $C$ fields define
a discrete Heisenberg group action $\text{H}(\mathbb{Z})$ which will play a central role
in the discussion of NS5-brane instanton effects in \S\ref{sec_NS5}. Eq. \eqref{heis0} completely
specifies the restriction of the $\IC^\times$-bundle
(or rather, its unit circle bundle $\mathscr{C}_{\sigma}$, with $S^1$-fiber
parametrized by the axion $\sigma$) in \eqref{IIAbundle} to the torus $\cT$.
The topology of the bundle over the full intermediate Jacobian $\cJ_W(X)$ will be discussed
in the next paragraph after discussing the one-loop corrected metric and in more detail in \S\ref{sec-topol}.

\subsubsection{Perturbative metric}
\label{sec_met}

At tree-level, the metric on $\cM_H$ belongs to the class of `semi-flat' QK metrics discovered
in \cite{Cecotti:1989qn,Ferrara:1989ik}. In particular, its restriction to
$\cM_C(X)$ is the special \kahler metric $g_{\cM_C(X)}$ deduced from the prepotential $F(X^\Lambda)$,
with \kahler potential \cite{Strominger:1990pd}
\be
\label{defcK}
\cK= -\log\left[\I \int_X \Omega^{3,0}\wedge \overline{\Omega^{3,0}} \right] =
-\log\left[ \I ( \bar X^\Lambda F_\Lambda - X^\Lambda \bar F_\Lambda ) \right] ,
\ee
while along $\cT\ltimes S^1$ it has continuous isometries.
As shown in  \cite{Gunther:1998sc, Antoniadis:2003sw,Robles-Llana:2006ez,Alexandrov:2007ec}, the one-loop
correction takes the metric outside the above class, still preserving flatness along $\cT\ltimes S^1$.
The resulting metric on $\cM_H$ can be written as
\be
g_{\text{pert}}= \frac{r+2c}{r^2(r+c)}\,\text{d}r^2+\frac{4(r+c)}{r}\,g_{\cM_C(X)}+
\frac{1}{r}\, g_{\cT}(c)+\frac{r+c}{16r^2(r+2c)}\, \big(\text{d}\sigma + \cA(c)\big)^2,
\label{pertmetricIIA}
\ee
where $r=e^{\phi}$ and the parameter $c=-\chi(X)/(192\pi)$ encodes the one-loop correction, governed solely by
the Euler number of $X$. Here $g_\cT(c)$ denotes a
deformation of the standard Weil metric on the torus $\cT$
which can be found in \cite{Alexandrov:2007ec}.
Most importantly, the connection $\cA(c)$ on the circle
bundle $\mathscr{C}_\sigma$ is given by
\be
\label{A1loop}
\cA(c)=  \tzeta_\Lambda \de \zeta^\Lambda -  \zeta^\Lambda \de \tzeta_\Lambda
+  8c \cA_K, \qquad
\cA_K=\frac{\I}{2}( \pa_{z^a} \cK \de z^a -  \pa_{z^{\ba}} \cK \de \bar z^{\ba})  .
\ee
Here $\cA_K$ is the K\"ahler connection on the Hodge bundle $\cL_X$
with  K\"ahler potential \eqref{defcK}. The  second  term in \eqref{A1loop}
follows by reducing the
topological coupling $B\wedge I_8$ in the ten-dimensional type IIA action and dualizing $B$ into $\sigma$
(see \cite{Alexandrov:2010ca} for details).
The tree-level metric is recovered by setting $c=0$.

The connection \eqref{A1loop}
implies that the circle bundle $\mathscr{C}_\sigma$ has non-trivial curvature both along the torus $\cT$,
in accordance with \eqref{heis0}, but also along the base $\cM_{C}(X)$
of the intermediate Jacobian $\cJ_W(X)$; the first Chern class is given by
\be
\label{Chcl}
c_1(\mathscr{C}_\sigma) = \omega_{\cT} + \frac{\chi(X)}{24}\,  \omega_{C},
\qquad
\omega_\cT \equiv \text{d}\tilde\zeta_\Lambda\wedge \text{d}\zeta^{\Lambda},
\qquad
\omega_{C}\equiv -\frac{1}{2\pi}\,\text{d}\cA_K .
\ee
We return to the topology of the axion circle bundle in relation to NS5-instantons in \S\ref{sec_NS5}.
For now, notice that putative higher loop corrections would in general induce
corrections to $c_1(\mathscr{C}_\sigma)$ suppressed by inverse powers of $r$, contradicting
the requirement that $c_1(\mathscr{C}_\sigma)\in H^2(\cM_{H},\IZ)$.
Note that for $\chi(X)>0$, the metric \eqref{pertmetricIIA} has a curvature singularity
at $r=-2c$ (while $r=0$, $r=-c$ are coordinate singularities). This singularity is expected to be resolved
once the full set of non-perturbative corrections is included \cite{Alexandrov:2014sya}.

\subsubsection{Twistor space description}
\label{sec_twistor}

The most convenient way of describing a QK manifold $\cM$
is via its twistor space $\cZ$ \cite{MR664330}. Recall that a quaternion-K\"ahler manifold of
real dimension $4n$ has holonomy group contained in $USp(n)\times SU(2)\subset SO(4n)$.
In particular, it has a triplet of almost complex structures $\vec{J}$ (defined locally up to $SU(2)$ rotations)
satisfying the quaternion algebra, corresponding two-forms $\vec{w}$
and a globally defined closed 4-form $\vec{w}\wedge \vec{w}$. The $J_i$'s
are not integrable unless the scalar curvature of $\cM$ vanishes, in which case $\cM$ is
hyperk\"ahler. Nevertheless, it is possible to encode the geometry
of $\cM$ complex analytically, by passing to its twistor space $\cZ$,
the total space of a canonical $\mathbb{P}^1$-bundle over $\cM$.  $\cZ$ carries
a canonical complex contact structure, given by the kernel of the $\cO(2)$-twisted, (1,0)-form
\be
Dt=\text{d}t + p_+ -\I p_3 t +p_- t^2,
\label{Dt}
\ee
where $t$ is a stereographic coordinate on $\mathbb{P}^1$ and $(p_{\pm}=-\tfrac12(p_1\mp\I p_2), p_3)$ denotes
the $SU(2)$-part of the Levi-Civita connection on $\cM$.

Locally on an open patch $\cU_i\subset \cZ$ there exists a function $\Phi_{[i]}$,
the `contact potential', which is holomorphic along the twistor lines (i.e. the fibers
of $\cZ\longrightarrow \cM$) and such that the product
\be
\cX^{[i]}=-4\I\, e^{\Phi^{[i]}}\, Dt/t
\label{cont-oneform}
\ee
is a holomorphic (i.e. $\bar\pa$-closed) one-form. The  nowhere vanishing holomorphic top-form
$\cX \wedge (\de \cX)^n $
defines the complex contact structure on $\cZ$. Locally, by a complex--contact analogue of
the Darboux theorem, one can always choose complex coordinates
$(\xi^{\Lambda}_{[i]}, \tilde\xi_\Lambda^{[i]}, \alpha^{[i]})$ in  $\cU_i$ such that
the contact one-form \eqref{cont-oneform} takes the canonical form \cite{Neitzke:2007ke,Alexandrov:2008nk}
\be
\cX^{[i]}= \text{d}\alpha^{[i]}+ \xi_{[i]}^\Lambda \text{d}\txi_{\Lambda}^{[i]} .
\label{contact1form}
\ee
In what follows it will often be convenient to combine $\xi^{\Lambda}$ and $\tilde\xi_\Lambda$
into a symplectic vector $\Xi=(\xi^{\Lambda}, \tilde\xi_\Lambda)$, and to define a variant
$\tilde\alpha = -2 \alpha - \tilde\xi_\Lambda\xi^{\Lambda} $ of the
coordinate $\alpha$ such that
\be
\cX^{[i]}= -\frac12 \left( \text{d}\tilde\alpha^{[i]}+ \txi^{[i]}_\Lambda \text{d}\xi^{\Lambda}_{[i]}
-\xi_{[i]}^\Lambda \text{d}\txi_{\Lambda}^{[i]} \right)
= -\frac12 \left( \text{d}\tilde\alpha^{[i]}+ \langle \Xi^{[i]}, \de \Xi^{[i]} \rangle \right) .
\label{contact1alt}
\ee
The global complex contact structure on $\cZ$ is then encoded into the set of complex contact
transformations between overlapping Darboux coordinate systems on $\cU_i\cap \cU_j$.
A  convenient way of specifying these contact transformations is via  a set of
Hamilton generating functions $H^{[ij]}(\xi, \txi, \alpha) \in H^1(\cZ, \cO(2))$, as explained
in detail in \cite{Alexandrov:2008nk,Alexandrov:2008gh}. The QK metric can be reconstructed
by (i) parametrizing the twistor lines, i.e. expressing the complex Darboux coordinates $(\Xi,\alpha)$
in terms of the local coordinates $(t,x^\mu)$ on $\mathbb{P}^1\times \cM$;
(ii)  evaluating the contact one-form (\ref{contact1form}) and matching the result with \eqref{Dt}
so as to extract the $SU(2)$ connection $\vec p$;
(iii) computing the quaternionic 2-forms $\vec{w}$ via
$\text{d}\vec{p}+\tfrac{1}{2}\vec{p}\wedge \vec{p} = \tfrac{\nu}{2}\vec{w}$
($\nu$ is the constant curvature of $\cM_H$);
(iv) constructing the space of $(1,0)$-forms with respect to $J_3$, by expanding
the differentials $(\de\Xi, \de\alpha)$ around $t=0$, and finally, (v)  contracting
the K\"ahler form $w_3$ with the complex structure $J_3$.

In this framework, the perturbative metric (\ref{pertmetricIIA}) is captured by the following Darboux coordinates
in the patch $\cU_0=\mathbb{P}^1\backslash\{0,\infty\}$
\cite{Alexandrov:2008nk} (building on earlier work \cite{Rocek:2005ij,Neitzke:2007ke,
Alexandrov:2007ec}):
\bea
\xi^{\Lambda} & =& \zeta^{\Lambda} +
2\sqrt{r+c} \, e^{\cK/2} \big(t^{-1}ÊX^{\Lambda}-t \, \bar{X}^{\Lambda}\big),
\nn \\
\tilde\xi_\Lambda &=& \tilde\zeta_\Lambda +
2\sqrt{r+c} \, e^{\cK/2} \big(t^{-1} F_\Lambda - t \, \bar{F}_\Lambda\big),
\label{Darbouxpert}\\
\tilde{\alpha} &=& \sigma +2\sqrt{r+c} \, e^{\cK/2} \big(t^{-1} W- t \,\bar{W}\big)-8\I \, c \log t,
\nn
\eea
where  $W \equiv F_\Lambda \zeta^{\Lambda}- X^{\Lambda}\tilde\zeta_\Lambda$,
whereas the contact potential coincides with the dilaton, $\Phi=\phi$.
The last term in the expression for $\tilde\alpha$ is the sole effect of the one-loop correction
in this framework (except for a field redefinition $r\to r+c$). Under a holomorphic rescaling
$\Omega^{3,0}\to e^{f}\Omega^{3,0}$, the \kahler potential $\cK$ and coordinates $t,\sigma$
vary according to $\cK\to \cK-f-\bar f$, $t\mapsto e^{\I\,\Im f} t$, $\sigma\mapsto \sigma - 8c\, \Im f$,
leaving \eqref{Darbouxpert} invariant.

For our purposes, it is important to note two key properties of the twistorial approach. First, quaternionic
isometries of  $\cM$ (i.e. preserving the 4-form $\vec{w}\wedge \vec{w}$) are classified by
the Cech cohomology group $H^0(\cZ,\cO(2))$ via the moment map construction, and therefore
lift to holomorphic actions on  $\cZ$ \cite{MR872143}.
In particular, the action of the Heisenberg group (\ref{heis0})
on $\cM_H$ lifts to a holomorphic action  on the Darboux coordinates \eqref{Darbouxpert} as
\be
\label{heis0twistor}
\bigl(\Xi,\tilde\alpha\bigr)\longmapsto
\bigl(\Xi+H ,\
\tilde\alpha + 2 \kappa +\left<\Xi, H \right> +2c(H)\bigr) .
\ee
The second property is that linear deformations of a QK space $\cM$ are classified by sections of
$H^1(\cZ,\cO(2))$ \cite{lebrun1994srp}.

\subsection{Type IIB}
\label{subsec_IIB}

We now turn to the perturbative HM moduli space $\hat\cM_H(\hat{X})$ in type IIB string theory
compactified on a CY threefold $\hat X$. Mirror symmetry requires that it should be isometric
to the previously discussed type IIA HM moduli space $\cM_H(X)$ whenever
$(X,\hat X)$ form a dual pair. The two spaces however come with different natural coordinates,
and it is important to determine the `mirror map' between the two sides.

On the type IIB side, the HM moduli space has a similar fibration structure as in \eqref{IIAbundle},
\be
\mathbb{C}^{\times} \, \longrightarrow  \, \hat\cM_H(\hat{X}) \, \longrightarrow\, \cJ_K(\hat{X})
\, \longrightarrow  \, \cM_K(\hat X) ,
\label{IIBfibration}
\ee
where the `even Jacobian' $\cJ_K(\hat{X})$ is a torus bundle over
$\cM_K(\hat X)$, the moduli space of complexified \kahler structures  on $\hat X$,
with fiber $\hat\cT=H^{\text{even}}(\hat{X}, \mathbb{R})/\hat \Gamma$ where $\hat\Gamma$
is a lattice which will be  specified below. Physically,
the $\mathbb{C}^{\times}$-fiber is parametrized by the type IIB dilaton $\tau_2=1/g_s$
and the NS-axion $\psi$, while $\hat\cT$ corresponds to the periods of
the  ten-dimensional RR form $C^{\text{even}}=C^{(0)}+C^{(2)}+C^{(4)}+C^{(6)}=H^{\rm even}(\hat X, \IR)$.
A convenient set of coordinates is given by~\cite{Louis:2002ny,Alexandrov:2008gh}
\be
\begin{split}
c^{0}&\,=C^{(0)},
\qquad
c^{a}=\int_{\gamma^{a}} C^{(2)},
\qquad
\cla=-\int_{\gamma_a} \left(C^{(4)}-\tfrac{1}{2}B\wedge C^{(2)}\right),
\\
\cl0&\, =-\int_{\hat{X}} \left(C^{(6)}-B\wedge C^{(4)}+ \tfrac{1}{3} B\wedge B\wedge C^{(2)}\right),
\qquad b^{a}+\I t^{a}=\int_{\gamma^{a}} (B+\I J).
\end{split}
\ee
where $\gamma_a, \, a=1,\dots, h^{1,1}, \,$Ê is a basis of 4-cycles in $H_4(\hat{X}, \mathbb{R})$,
and $\gamma^{a}$ is the dual basis of 2-cycles in $H_2(\hat{X}, \mathbb{R})$.

By classical mirror symmetry, $\cM_K(\hat X)=\cM_C(X)$, for a suitable map between the complex structure moduli
$z^a=X^a/X^0$ and the \kahler moduli $(b^a, t^a)$. In the large volume limit on the type IIB side,
the prepotential on $\cM_K(\hat X)$ is given by
\be
F^{\rm cl}=-\kappa_{abc}\, \frac{X^a X^b X^c}{6X^0}+\hf\, A_{\Lambda\Sigma}X^\Lambda X^\Sigma\, ,
\label{classprep}
\ee
where $\kappa_{abc}$ is the triple intersection product on $H^2(\hat X,\IZ)$ and $A_{\Lambda\Sigma}$ is a real
symmetric matrix which does not affect the K\"ahler potential, but which is important for
consistency with charge quantization \cite{Alexandrov:2010ca}. In this limit, the mirror map reduces to $z^a=b^a+\I t^a$
for a suitable choice of symplectic basis on the type IIA side adapted
to the point of maximal unipotent monodromy.
The classical metric on $\hat\cM_H(\hat{X})$ then takes the semi-flat form
\eqref{pertmetricIIA} with $c=0$ and the prepotential \eqref{classprep},
provided $\cM_K(\hat X)$ is identified with $\cM_C(X)$ and the
natural coordinates on the type IIB side are related to the coordinates \eqref{defC} on the
type IIA side by  \cite{Bohm:1999uk}:
\be
\label{symptobd}
\begin{split}
r &=  \frac{ \tau_2^2}{2} \, \cV\, ,
\qquad
X^{a}/X^{0}=z^a=b^a+\I t^a\, ,
\qquad
\zeta^0=\tau_1\, ,
\qquad
\zeta^a = - (c^a - \tau_1 b^a)\, ,
\\
\tzeta'_a &=  \cla+ \frac{1}{2}\, \kappa_{abc} \,b^b (c^c - \tau_1 b^c)\, ,
\qquad
\tzeta'_0 =\, \cl0-\frac{1}{6}\, \kappa_{abc} \,b^a b^b (c^c-\tau_1 b^c)\, ,
\\
\sigma &= -2 (\psi+\frac12  \tau_1 \cl0) + \cla (c^a - \tau_1 b^a)
-\frac{1}{6}\,\kappa_{abc} \, b^a c^b (c^c - \tau_1 b^c)\, ,
\end{split}
\ee
where $\cV=\frac16\, \kappa_{abc}t^at^bt^c$ denotes the volume of  $\hat{X}$ and the prime denotes fields obtained by
the symplectic transformation removing the quadratic term in \eqref{classprep},
namely,
\be
\tzeta'_\Lambda=\tzeta_\Lambda-A_{\Lambda\Sigma}\zeta^\Sigma.
\label{symplec-tr}
\ee
By mirror symmetry, the lattice\footnote{More precisely,
$\Gamma$ and $\hat\Gamma$ are local system of lattices over $\cM_C(X)$ and
$\cM_K(\hat X)$,  due to monodromies.}
$\hat\Gamma\subset H^{\text{even}}(\hat{X}, \mathbb{R})$ must be (indeed, is)
the image of the lattice $\Gamma\subset H^3(X,\IR)$ under the map \eqref{symptobd}
between the type IIA RR fields $\zeta^\Lambda,\tzeta_\Lambda$ and the type IIB RR fields $c^0,c^a,\tilde c_a,\tilde c_0$.
Beyond the large volume limit, the prepotential \eqref{classprep} and mirror map \eqref{symptobd}
acquire worldsheet instanton corrections, governed by the genus zero Gopakumar-Vafa invariants
$n^{(0)}_{q_a}$ of $\hat X$ \cite{RoblesLlana:2006is,Alexandrov:2008gh,Alexandrov:2009qq,Alexandrov:2012bu}.
Together with the one-loop correction proportional to
$\chi(\hat X)$, this produces the same metric \eqref{pertmetricIIA} as on the type IIA side.

Besides establishing mirror symmetry, the mirror map \eqref{symptobd} has another virtue:
it exposes the invariance of the HM moduli space $\cM_H(X)=\hat\cM_H(\hat{X})$,
in the large volume/weak coupling limit where it holds, under the action of $SL(2,\IR)$, corresponding
to the continuous S-duality symmetry of ten-dimensional type IIB supergravity.
Of course, this continuous symmetry is broken by quantum corrections, but there is
overwhelming evidence that a discrete $SL(2,\IZ)$ subgroup remains unbroken, providing
a strong constraint on possible non-perturbative effects  \cite{Green:1997tv}. The action
of $g=(\begin{smallmatrix} a & b \\ c & d  \end{smallmatrix})\in SL(2,\mathbb{Z})$
 is simplest in type IIB variables \cite{Gunther:1998sc,Bohm:1999uk}:
\be\label{SL2Z}
\begin{split}
&\quad \tau \mapsto \frac{a \tau +b}{c \tau + d} \, ,
\qquad
t^a \mapsto t^a |c\tau+d| \, ,
\qquad
\cla\mapsto \cla - c_{2,a} \varepsilon(g)\, ,
\\
&
\begin{pmatrix} c^a \\ b^a \end{pmatrix} \mapsto
\begin{pmatrix} a & b \\ c & d  \end{pmatrix}
\begin{pmatrix} c^a \\ b^a \end{pmatrix},
\qquad
\begin{pmatrix} \cl0 \\ \psi \end{pmatrix} \mapsto
\begin{pmatrix} d & -c \\ -b & a  \end{pmatrix}
\begin{pmatrix} \cl0 \\ \psi \end{pmatrix} .
\end{split}
\ee
In  this action, we have included a shift of the RR coordinate $\cla$,
overlooked in early studies but crucial for maintaining S-duality invariance under D3
and D5-NS5 instanton corrections \cite{Alexandrov:2010ca,Alexandrov:2012au}, as we shall
see in  \S\ref{sec_D3} and \S\ref{sec_NS5}. Here, $c_{2,a}$ and $\varepsilon(g)\in \mathbb{Q}$
are defined by
\be
c_{2,a}\equiv \int_{\gamma_a} c_2(\hat{X}) ,
\qquad
\eta\left(\frac{a\tau+b}{c\tau+d}\right)/\eta(\tau)=\expe{ \varepsilon(g)} (c\tau+d)^{-1/2},
\label{multeta}
\ee
where $\eta(\tau)$ is the Dedekind eta-function and $\expe{x}=\exp(2\pi\I\, x)$.
We stress that as defined so far, the metric on
$\hat\cM_H(\hat{X})$ is only invariant under $SL(2,\IZ)$ in the strict infinite volume,
zero string coupling limit, where it is actually enhanced to $SL(2,\IR)$. Both worldsheet instanton
corrections to the classical prepotential \eqref{classprep} and the one-loop correction
break this symmetry, and it is necessary to include non-perturbative effects in order to recover it.
The holomorphic action of S-duality in twistor space will be described in \S\ref{subsec-Stw}.

\section{D-instantons, wall-crossing and the QK/HK correspondence}
\label{sec_Dinst}

The perturbative metric (\ref{pertmetricIIA}), while being valid to all orders at small string coupling
$g_s$, is expected to receive non-perturbative corrections of order $e^{-1/g_s}$,
due to Euclidean D-branes wrapping supersymmetric cycles in $X$  (or $\hat X$).
In this section we discuss some general aspects of D-instantons and their relation with
Donaldson-Thomas (DT) invariants,  describe how they modify the twistorial description of \S\ref{sec_twistor},
and how they result in a smooth quantum corrected metric, despite
discontinuities of the DT invariants across certain walls in complex
structure (or \kahler) moduli space.
For this purpose, a new duality between quaternion-K\"ahler and
hyperk\"ahler manifolds  will turn out to be useful.

\subsection{D-instantons, Donaldson-Thomas invariants and wall-crossing}
\label{subsec-DTD}

\subsubsection{Derived category of D-instantons}
\label{DTD}

On the type IIA side, the leading corrections to the perturbative hypermultiplet metric
$g_{\text{pert}}$ described in \S\ref{sec_met} come from  Euclidean D2-branes wrapping
Lagrangian 3-cycles (sLags) in $X$ endowed with a flat $U(1)$ connection.
On the type IIB side, they correspond to superpositions of  D(-1)-D1-D3-D5 instantons wrapping
complex even-dimensional cycles, or more generally  coherent sheaves  (holomorphic vector bundles supported
on (singular) submanifolds). Most generally, D-instantons are
objects in a bounded derived Fukaya category  ${\rm D^{b}Fuk}(X)$ on the type IIA side,
or the derived category of coherent sheaves ${\rm D^{b}Coh}(\hat X)$ on the type IIB side \cite{Sharpe:1999qz,Douglas:2000ah}.
Each of them is graded by the Grothendieck group, an extension
of the lattice $\Gamma=H_3(X, \mathbb{Z})$ or
$\hat\Gamma\subset H_{\rm even}(\hat X, \mathbb{Z})$ of electromagnetic charges.
The fact that the same categories
also govern the spectrum of BPS states in type IIB on $X$ and type IIA on $\hat X$, respectively,
can be understood by compactifying on a circle down to 3 space-time dimensions: T-duality along
the circle exchanges 4-dimensional D-instantons with 4-dimensional BPS states whose worldline winds
around the circle \cite{Alexandrov:2008gh}. Kontsevich's homological mirror symmetry conjecture \cite{MR1403918},
the mathematical counterpart of non-perturbative mirror symmetry \cite{Ferrara:1995yx},
asserts that these two categories are isomorphic when $(X,\hat X)$ is a dual pair, in particular
$\Gamma\simeq \hat\Gamma$.

\subsubsection{Stability and DT invariants}
\label{stabDT}

Among all the objects in the derived category, those which correspond to supersymmetric,
elementary D-instantons (or dually, one-particle BPS states) are the
semi-stable ones  \cite{Douglas:2000ah}. Stability
can be assessed using the central charge $Z$, a homomorphism $Z\, :\, \Gamma \to
\mathbb{C}$ which varies holomorphically over  $\cB$ ($=\cM_C(X)$ or $\cM_K(\hat X)$)
and which determines the classical action (or dually, the mass) of the instanton.
In type IIA we have
\be
Z_\gamma(z)=e^{\cK/2}\int_\gamma \Omega^{3,0},
\label{centralcharge}
\ee
while in the large-volume limit in type IIB
\be
\label{cchargeIIB}
Z_\gamma(z)=\int_{\hat{X}} e^{B+\I J} \text{ch}({\mathscr{E}}) \sqrt{\text{Td}(\hat{X})},
\ee
where $\mathscr{E}$ is a coherent sheaf, whose Mukai vector
$\text{ch}({\mathscr{E}}) \sqrt{\text{Td}(\hat{X})}\in
H^{\text{even}}(\hat{X},\mathbb{Q})$ is identified with the charge
vector $\gamma$ of $E$. Semi-stability is most easily defined for Abelian categories as
follows.
An object $F$ with charge $\gamma$ is called semi-stable if for every subobject
$F'\subset F$ with charge $\gamma'$, $\varphi(\gamma')\leq \varphi(\gamma)$,
where $\varphi(\gamma)$ is the argument of the central charge $Z_\gamma(z)$.
$F$ is called stable if the inequality is strict for strict subobjects.
This notion can be extended  to derived categories, by considering an Abelian
subcategory of the derived category, the ``heart of
the $t$-structure'' \cite{DouglasTopMirrorMap,MR2373143}.
On the IIA side, semi-stable objects of ${\rm D^bFuk}(X)$ are {\it special} (or calibrated)
Lagrangian
cycles $L$, i.e. such that the phase of
$\Omega\vert_L/\de V_L$ is constant, where $\de V_L$ is the volume form on
$L$  \cite{MR666108}.  On the IIB side in the infinite volume limit, semi-stable objects are the semi-stable coherent
sheaves in the classical sense of Gieseker stability.

An important property of semi-stable objects is that their space of deformations is finite-dimensional,
although it can be singular. The generalized DT invariant $\Omega(\gamma;z)$ is defined as the (weighted)
Euler number of this moduli space. It is the mathematical
counterpart of the BPS index, which counts BPS black holes or instantons
of charge $\gamma$. It is a locally constant function of the moduli $z$
(through the central charge $Z_\gamma$),
away from certain walls of marginal stability described below.
It is also monodromy invariant, in the sense that $\Omega(M\cdot \gamma;M\cdot z)=\Omega(\gamma;z)$,
where $M\in Sp(m;\mathbb{Z})$ is the symplectic rotation
induced by a monodromy along  a loop in $\cB$.
Homological mirror symmetry implies that the DT invariants $\Omega(\gamma;z)$ associated to
${\rm D^{b}Fuk}(X)$ and ${\rm D^{b}Coh}(\hat X)$ are the same, provided the charges
$\gamma$ and moduli $z$ are related according to the classical mirror map. As we shall see,
this guarantees that the D-instanton corrected HM moduli spaces $\cM_H(X)$ and $\hat{\cM}_{ H}(\hat X)$ are isometric.

\subsubsection{Wall-crossing}
Physically, the jump of the
DT invariants $\Omega(\gamma,z)$ across codimension one walls  in $\cB$ corresponds
to the decay of bound states into more elementary stable constituents. For any pair
of charge vectors $(\gamma_1,\gamma_2)$, the decay of a
D-brane of charge $\gamma = M\gamma_1+N\gamma_2$ into constituents of charges
$M_i\gamma_1+N_i\gamma_2$  with $\sum (M_i,N_i)=(M,N)$
is energetically possible only if the phases of the central charges align,
i.e. $\varphi(\gamma_1) = \varphi(\gamma_2)$, which defines
the wall of marginal stability $W(\gamma_1, \gamma_2)\subset \cB$. Let
$z_{\pm}\in \cB$ denote two points infinitesimally displaced on either side of such a wall.
We can always choose the basis $\gamma_1,\gamma_2$ of the two-dimensional lattice $\IZ\gamma_1+\IZ\gamma_2$
such that only the first and third quadrants are populated on either side of the
wall, $\Omega^\pm(M\gamma_1+N\gamma_2)=0$ if $MN\leq 0$ \cite{Andriyash:2010qv}.

Several formulae exist in the mathematics and physics literature for how to
compute the jump of $\Omega(M\gamma_1+N\gamma_2;z)$  across the wall
$W(\gamma_1, \gamma_2)$ (see, e.g., \cite{Pioline:2011gf} for a review). The one relevant here
is the Kontsevich-Soibelman (KS) formula \cite{ks}, which has a clear geometric interpretation.
To write their formula, one introduces the Lie algebra of (twisted)
infinitesimal symplectomorphisms of
the complex torus $\Gamma\otimes_{\mathbb{Z}} \mathbb{C}^{\times}$ generated by
vector fields $(e_\gamma)_{\gamma\in \Gamma}$ satisfying
\be
[e_{\gamma}, e_{\gamma'}]=(-1)^{\left<\gamma, \gamma'\right>} \left<\gamma, \gamma'\right> e_{\gamma+\gamma'}.
\ee
For any $\gamma\in\Gamma$, $z\in\cB$
we also define the group element
\be
U_\gamma(z) = \exp\left( \Omega(\gamma;z)\, \sum_{n=1}^{\infty} \frac{e_{n\gamma}}{n^2}\right).
\label{KSgroupelement}
\ee
The KS wall-crossing formula then asserts the following equality between
oppositely ordered infinite products  of symplectomorphisms \cite{ks}
\be
\prod_{\gamma=m\gamma_1+n\gamma_2\atop m \geq 0, n\geq 0}^{\curvearrowleft} U_\gamma(z_+)
=\prod_{\gamma=m\gamma_1+n\gamma_2\atop m\geq 0, n\geq 0}^{\curvearrowright} U_\gamma(z_-).
\label{KSWCF}
\ee
By projecting this equality on finite dimensional quotients, one can
determine $\Delta\Omega=\Omega(\gamma;z_+)-\Omega(\gamma;z_-)$ for any
$\gamma = M\gamma_1+N\gamma_2$. This formula was interpreted physically
in the context of $\cN=2$ gauge theory \`a la Seiberg-Witten in \cite{Gaiotto:2008cd}, as ensuring
the smoothness of the hyperk\"ahler metric on the Coulomb branch
of the gauge theory on $\mathbb{R}^3\times S^1$. From this point of view, the $U_\gamma$'s
are complex symplectomorphisms relating different Darboux coordinate systems on
the twistor space.  Below we will show that an extension of \eqref{KSWCF},
where the $U_\gamma$'s  now are complex contact transformations, also ensures the
smoothness of the HM moduli space $\cM_H$ in $\cN=2$ string vacua across walls in $\cB$.

\subsection{D-instantons in twistor space}
\label{sec_DinstTwistor}

Away from the zero-coupling limit, the perturbative hypermultiplet metric
(\ref{pertmetricIIA}) receives non-perturbative corrections due to D-brane instantons.
In the one-instanton approximation these corrections take the schematic
form \cite{Alexandrov:2008gh}
\be
g_{D} \sim \sum_{\gamma\in \Gamma}\qrD\, \bar{\Omega}(\gamma;z)
\exp\Big(-8\pi \sqrt{r} \,|Z_\gamma|  -2\pi \I \left<\gamma, C\right>\Big) ,
\label{Dinstseries}
\ee
where the exponential is the classical action of the D-instanton, with $\sqrt{r}\sim1/g_s$. The prefactor in principle
originates from integrating the fluctuation determinant around the classical solution
over collective coordinates. It is natural to expect that it is proportional to the DT invariant
$\Omega(\gamma,z)$ introduced in \S\ref{stabDT}, however consistency with wall-crossing
will dictate the less obvious product of a quadratic refinement $\qrD$,
analogous to the one in \eqref{shiftcoc}, with the `rational DT invariant' \cite{ks,Joyce:2008pc,Manschot:2010qz}
\be
\bOm(\gamma;z) =\sum_{d|\gamma}
\frac{\Omega(\gamma/d;z)}{d^2}.
\label{ratDT}
\ee
When $\gamma$ is a primitive charge vector, $\bOm(\gamma;z) = \Omega(\gamma;z)$.

In order to incorporate these corrections to the metric while maintaining its quaternion-K\"ahler structure,
it is best to do this at the level of the twistor space $\cZ$.
As explained in \cite{Alexandrov:2008gh,Alexandrov:2009zh,Alexandrov:2011ac},
in close analogy with the field theory construction in \cite{Gaiotto:2008cd},
the instanton corrections modify the contact structure
on $\cZ$, presented in \S\ref{sec_twistor}, by replacing the patch $\cU_0$ around
the equator of $\mathbb{P}^1$ with an infinite set of angular sectors separated by so called  ``BPS-rays''
\be
\ell_\gamma =\{ t\in \mathbb{P}^1 \ :\  Z_\gamma(z) \, t^{-1} \in \I\mathbb{R}_-\} ,
\label{QKBPSray}
\ee
where $Z_\gamma(z)$ is the central charge function \eqref{centralcharge}.
Across $\ell_\gamma$ the  Darboux coordinates $(\xi^{\Lambda}, \tilde\xi_\Lambda, \tilde\alpha)$
must jump by a complex contact transformation. We postulate that the jump of
the holomorphic Fourier modes on the torus
$\Gamma \otimes_{\mathbb{Z}} \mathbb{C}^{\times}$,
\be
\cX_\gamma=\expe{ - \left<\gamma, \Xi\right>}=e^{-2\pi\I\(q_\Lambda\xi^\Lambda-p^\Lambda\txi_\Lambda\)},
\label{holfourier}
\ee
is the standard KS symplectomorphism
\be
U_\gamma \, :\,  \cX_{\gamma'} \, \longmapsto\,
 \cX_{\gamma'}\left(1-\qrD\cX_{\gamma}\right)^{\Omega(\gamma)\left<\gamma, \gamma'\right>} .
\label{KStrans}
\ee
Requiring that the contact one-form \eqref{contact1form} is preserved determines
the discontinuity in the remaining Darboux coordinate $\tilde\alpha$. As a result, the full
contact transformation is given by
\be
V_\gamma \, :\, ( \cX_{\gamma'},\, \tilde\alpha) \, \longmapsto\,
\left( \cX_{\gamma'}\left(1-\qrD\cX_{\gamma}\right)^{\Omega(\gamma)\left<\gamma, \gamma'\right>},\,
\tilde\alpha + \frac{\Omega(\gamma)}{2\pi^2}\,
L_{\qrD}\left(\qrD\cX_{\gamma}\right)
\right) ,
\label{KStrans2}
\ee
where $L_{\epsilon}(x)$ is a variant of the Rogers dilogarithm,
\be
L_{\epsilon}(x)\equiv \text{Li}_2(x)+\tfrac{1}{2}\log(\epsilon^{-1}x)\log(1-x) .
\ee

Requiring further that the Darboux coordinates reduce to the uncorrected ones \eqref{Darbouxpert}
near $t=0$ and $t=\infty$,  one may recast the gluing conditions  \eqref{KStrans2}
across the BPS rays $\ell_\gamma$ as a system of integral
equations for the Fourier modes $\cX_\gamma$ \cite{Gaiotto:2008cd,Alexandrov:2009zh},
\be
\cX_\gamma(t) = \cX_\gamma^{\text{sf}}(t) \, \exp\left[\frac{1}{4\pi \I}\sum_{\gamma'}
\Omega(\gamma')\left<\gamma, \gamma'\right>\int_{\ell_{\gamma'}}\frac{\text{d}t'}{t'}
\frac{t+t'}{t-t'} \log\left(1-\qrDp\cX_{\gamma'}(t')\right)\right],
\label{TBA}
\ee
where $\cX^{\text{sf}}_\gamma$ are the `semi-flat' Fourier modes obtained from  \eqref{Darbouxpert}
\be
\cX^{\text{sf}}_\gamma(t)= \exp\left[-2\pi \I \left(\left<\gamma, C\right>
+\frac{\tau_2}{2}\, e^{-\cK/2} \(t^{-1} Z_\gamma - t \bar{Z}_\gamma\)\right)\right].
\label{Xsf}
\ee
The remaining coordinate $\tilde\alpha$ and the contact potential, which in this case is globally defined,
independent of $t$ and can be identified with the 4-dimensional dilaton $\Phi=\phi$, are then obtained
from the solutions of  \eqref{TBA} via \cite{Alexandrov:2009zh,Alexandrov:2011ac}
\bea
\tilde\alpha\!\!\!\!&=&\sigma + t^{-1}\cW-t \bar{\cW}
+\frac{\I\chi(X)}{24\pi}\log t +\frac{\I}{8\pi^3}\sum_\gamma \Omega(\gamma)
\int_{\ell_\gamma} \frac{\text{d}t'}{t'}\, \frac{t+t'}{t-t'}\,L_{\qrD}\left(\qrD\cX_\gamma\right),
\nn \\
\\
e^{\phi} \!\!\!\!&=& \frac{\tau_2^2}{16}\, e^{-\cK}+\frac{\chi(X)}{192\pi}-\frac{\I\tau_2}{64\pi^2}\,e^{-\cK/2}\sum_\gamma \Omega(\gamma)
\int_{\ell_\gamma} \frac{\text{d}t}{t} \( t^{-1} Z_\gamma-t\bZ_\gamma\) \log\left(1-\qrD\cX_\gamma\right),
\nn \\
\label{contactpot-D} 
\eea
where the expression for $\cW$ can be found in \cite{Alexandrov:2011ac}.
While \eqref{TBA} cannot be solved exactly in general,  it can be solved approximately by
first plugging in $\cX_\gamma=\cX_\gamma^{\text{sf}}$ on the r.h.s., computing the l.h.s. and iterating.
This generates a formal infinite series of terms labelled by decorated rooted trees,
interpreted as multi-instanton corrections \cite{Gaiotto:2008cd,Stoppa:2011}.
The first term in this expansion,
known as the one-instanton approximation, is
an integral governed by a saddle point at $t=\I\sqrt{Z_\gamma/\bar Z_{\gamma}}$,
 which produces a result of the  expected form
 \eqref{Dinstseries}. Ref. \cite{Alexandrov:2014wca} argued that the
 third term on the right hand side in \eqref{contactpot-D} corresponds to the contribution
 of  multi-particle states to the Witten index (whereas
 $\Omega(\gamma)$ counts single-particle BPS states).

It should be stressed that the gluing conditions \eqref{KStrans2} apply only in an open set on
$\cZ$ away from any wall of marginal stability. Across such a wall, the DT invariants
$\Omega(\gamma)$ will jump, but so will the ordering of the BPS rays $\ell_\gamma$.
By the same reasoning as in \cite{Gaiotto:2008cd}, the consistency of the
construction and the smoothness of the instanton corrected metric (including all multi-instanton
corrections) requires an analogue
of the KS wall-crossing formula \eqref{KSWCF}, where the  $U_\gamma$'s are replaced by
$V_\gamma$'s, and equality holds modulo the axion periodicity $\tilde\alpha\to \tilde\alpha+2\kappa$, $\kappa\in\IZ$.
On the other hand, unlike the HK situation in \cite{Gaiotto:2008cd},
the twistor space $\cZ$ is not a trivial fibration $\cM_H\to\cZ\to\CP$, but rather an opposite,
non-trivial  fibration $\CP\to\cZ\to\cM_H$, and  the above construction does not address
the global structure of the twistor space.
To circumvent this problem, it will be convenient to
relate the  twistor space of the D-instanton corrected QK manifold to the
twistor space of a `dual' HK manifold, using a
general correspondence between QK and HK manifolds with isometries, which we now explain.

\subsection{The QK/HK correspondence}
\label{sec_QKHK}

Let us first recall the notion of \emph{hyperholomorphic  line bundle} on a HK
manifold $\cM'$:   A  line bundle $\mathscr{L}\to \cM'$
is hyperholomorphic if  its first Chern class $c_1(\mathscr{L})$
is of type $(1,1)$ with respect to the whole $S^2$ of complex structures
on $\cM'$ \cite{MR967469,MR1139657,MR1486984}. In real dimension 4 this reduces to
the notion of self-dual curvature. A hyperholomorphic connection is
a one-form $\lambda$ whose  curvature $\de\lambda$ satisfies the same
condition. It is the curvature of a hyperholomorphic line bundle if and only if
$\de\lambda \in H^2(\cM', \mathbb{Z})$.

\subsubsection{Theorem \cite{Haydys,Alexandrov:2011ac,Hitchin:2012}}

Given a quaternion-K\"ahler manifold $\cM$ with a quaternionic circle action generated
by a Killing vector $\kappa$, there exists  a `dual' hyperk\"ahler manifold
$\cM'$ of the same dimension, equipped with a circle action generated by the Killing vector $\kappa'$,
that fixes one of the complex structures, $J'_3$, and rotates $J'_1, J'_2$. Choosing
coordinates $\theta$, $\theta'$ adapted to the circle actions on both sides, such that
$\kappa=\pa_{\theta}$, $\kappa'=\pa_{\theta'}$, the QK metric on $\cM$ and HK metric on $\cM'$ are related by
\be
\label{dsqkgen}
\de s^2_{\cM} =  \tau\(\de\theta+\Theta\)^2 + \de s^2_{\cM/\pa_\theta}, \qquad
\de s^2_{\cM'} =(\nu+\rho) \(\de\theta'+\Theta'\)^2 +\de s^2_{\cM'/\pa_{\theta'}}
\ee
with
\be
\de s^2_{\cM'/\pa_{\theta'}} =\frac{\de\rho^2}{\rho} + 4 \rho \, |p_+|^2   -2\rho \, \de s^2_{\cM/\pa_\theta}.
\ee
Here, $\rho$ is a function on $\cM'/\pa_{\theta'}$ defined as the moment map of $\kappa'$ with respect
to $J'_3$; it is identified with the function $1/(2|\vec\mu|)$ on $\cM/\pa_\theta$,
where $\vec\mu$ is quaternionic moment map of $\kappa$ on $\cM$. $\tau$ and $\nu$ are functions
on $\cM/\pa_\theta$ and $\cM'/\pa_{\theta'}$, respectively, related by $\tau=\frac{\nu+\rho}{2\rho^2\nu}$.
Finally, $p_+$ is the $+$ component of the $SU(2)$ connection on $\cM$, related to the same
component of the  $SU(2)$ connection on $\cM'$ via $p'_+=\rho\, e^{\I\theta'} p_+$, while
the one-forms $\Theta$ and $\Theta'$ appear in the decomposition of the third component,
$p_3=-\frac{1}{\rho}(\de\theta+\Theta)+\Theta'$, $p_3'=\rho(\de\theta'+\Theta')-\Theta$.

Under this correspondence, the HK manifold $\cM'$ is naturally
endowed with a hyperholomorphic connection\footnote{Hyperholomorphicity is guaranteed
by the second equation in \eqref{hyphol}.}
\be
\lambda = \nu\, (\de\theta' + \Theta') + \Theta,
\qquad
\de\lambda=2\I \, \pa\bar\pa\rho - w_3' \in H^2(\cM',\IZ),
\label{hyphol}
\ee
where $w_3'$ is the \kahler form on $\cM'$ associated to $J_3'$ and $\pa$ is the
Dolbeault derivative in the same complex structure. Conversely, given a HK manifold with
a hyperholomorphic line bundle, there exists a one-parameter family of QK metrics
given by the same formula. The one-parameter  ambiguity stems from the fact that the moment
map $\rho$ of $\kappa'$ is defined up to an additive constant, which can be absorbed in
a shift of $\nu$. This affects the hyperholomorphic connection $\lambda$ but not its curvature.
Examples of such dual pairs are provided by the rigid and (one-loop deformed)
local $c$-map spaces associated to the same prepotential $F(X)$
\cite{Alexandrov:2011ac,Alekseevsky:2012fu}.

\subsubsection{Twistor space realization}
\label{sec_QKHKtwistor}

The QK/HK correspondence is most easily understood by using Swann's relation between
QK manifolds and \hk cones \cite{MR1096180}. Indeed,
the total space $\cS$ of the $\cO(-2)$ line
bundle over the twistor space $\cZ$ of a QK manifold $\cM$ carries a canonical HK cone metric.
Any quaternionic circle action of $\cM$ lifts to a triholomorphic circle action on $\cS$.
Taking the  hyperk\"ahler quotient of $\cS$ with respect to this circle action then produces
the dual hyperk\"ahler manifold $\cM'$, equipped with a natural hyperholomorphic
circle bundle  \cite{MR967469,MR1139657}.

For our purposes it will be more useful to realize the QK/HK correspondence directly at the
level of the twistor spaces $\cZ$ and $\cZ'$, without invoking the Swann bundle. For this purpose,
choose local contact Darboux coordinates $(\xi^{\Lambda}, \tilde\xi_\Lambda, \alpha)$ on the QK side,
such that the Killing vector $\kappa$ globally lifts to  $\pa_{\alpha}$ (the Reeb vector for the
contact one-form \eqref{cont-oneform}).
This implies that contact transformations between different patches must reduce to symplectomorphisms
of $(\xi^{\Lambda}, \tilde\xi_\Lambda)$, supplemented by a suitable,
$\Xi$-dependent shift of $\alpha$. On the HK side, we choose
local (symplectic) Darboux coordinates $(\eta^{\Lambda}, \mu_\Lambda)$ on $\cZ'$ such that the holomorphic
symplectic form on $\cZ'$ is $\text{d}\eta^{\Lambda} \wedge \text{d}\mu_\Lambda$.
We further choose coordinates $x^\mu$ on $\cM'/\pa_{\theta'}$, $\zeta$
on the $\CP$ base on the HK side and $t$ on the $\CP$ fiber on the QK side,
such  that $\zeta=0,\infty$ correspond to the complex structure $J_3'$ preserved by $\kappa'$.
The fact that $\kappa'$ rotates $J_1'$ into $J_2'$ means that the Darboux coordinates
$(\eta^{\Lambda}, \mu_\Lambda)$ on $\cZ'$ depend only on $x^\mu$ and  $\zeta e^{-\I\theta'}$.
Moreover, it implies that transition functions between different patches must be complex symplectomorphisms
of $(\eta^{\Lambda}, \mu_\Lambda)$, independent of $\zeta$.
Then the correspondence shows that the Darboux coordinates
$(\xi^{\Lambda}, \tilde\xi_\Lambda)$ on $\cZ$, as functions of $(x^\mu,\theta,t)$, can be identified
with $(\eta^{\Lambda}, \mu_\Lambda)$ for $t=\zeta e^{-\I\theta'}$. In particular, the complex contact
structure on $\cZ$ and symplectic structure on $\cZ'$ can be described globally by the same
symplectomorphisms, supplemented on the QK side by a suitable shift of $\alpha$. In fact, the
Darboux coordinate $\alpha$ provides, on the dual HK side, a holomorphic section
\be
\Upsilon^{[i]}\equiv e^{2\pi i \alpha^{[i]}}
\label{Ups}
\ee
of a line bundle $\mathscr{L}_{\cZ}'$ over $\cZ'$, with holomorphic connection given by the
contact one-form \eqref{contact1form}. This section is non-zero along each twistor line, and
hence by the Atiyah-Ward twistor correspondence (see \cite{ward-wells}) yields a
hyperholomorphic line bundle $\mathscr{L}$ over $\cM'$ with connection
\be
\lambda = \frac{1}{4} \left(\bar{\partial}^{(\zeta)} \alpha+\partial^{(\zeta)}\bar{\alpha}\right),
\label{hypercon}
\ee
which can be shown to agree with \eqref{hyphol}. It is also worth noting that the contact
potential $e^\phi$ on the QK side is identified with the moment map $\rho$ on the HK side.

\subsection{Wall-crossing revisited}
\label{wallD}

After this digression on the QK/HK correspondence, we now return to the problem of
wall-crossing on the D-instanton corrected  HM moduli space $\cM_H$.
Since D-instanton corrections are independent of the NS-axion $\sigma$, they preserve the
quaternionic Killing vector $\partial_\sigma$. Therefore, by {the QK/HK correspondence,
we can trade the construction of the QK space $\cM_H$ with that of a HK space $\cM'_{H}$ equipped
with a hyperholomorphic connection $\lambda$. Twistorially, this is equivalent to constructing
the twistor space $\cZ'$ and complex line bundle $\mathscr{L}_{\cZ'}$.
The space $\cZ'$, parametrized by the Darboux coordinates $(\xi^{\Lambda}, \txi_\Lambda)$,
is defined by the same gluing conditions \eqref{KStrans} as in \cite{Gaiotto:2008cd},
specialized to the case where the prepotential is homogeneous.
To construct\footnote{The construction of a canonical
hyperholomorphic connection on the Coulomb branch of $\cN=2$ gauge theories
was independently given in \cite{Neitzke:2011za}.}
$\mathscr{L}_{\cZ'}$,
one must lift the symplectomorphisms $U_\gamma$ to complex gauge transformations $V_\gamma$
preserving  the holomorphic connection \eqref{contact1form}, which we identify with
the contact transformations \eqref{KStrans2}.
The consistency of these gluing
conditions across walls of marginal stability require that the KS formula (\ref{KSWCF}) lifts to
\be
\prod_{\gamma=m\gamma_1+n\gamma_2\atop m\geq 0, n\geq 0}^{\curvearrowleft}V_\gamma(z_+)
=\prod_{\gamma=m\gamma_1+n\gamma_2\atop m\geq 0, n\geq 0}^{\curvearrowright} V_\gamma(z_-).
\label{contactWCF}
\ee
Clearly, assuming \eqref{KSWCF} is satisfied, \eqref{contactWCF} could  fail at most by
a translation $\tilde\alpha \to\tilde\alpha + \Delta\tilde\alphaÊ$
along the $\mathbb{C}^{\times}$-fiber. The global existence of $\mathscr{L}_{\cZ'}$ requires
$\Delta\tilde\alpha\in 2\IZ$, the natural ambiguity in the coordinate $\tilde \alpha$.
To see why this is so, let us rewrite  (\ref{contactWCF})  as an identity $\prod_s V_{\gamma_s}^{\epsilon_s}=1$
by assembling all operators on one side. Here, $\epsilon_s$ is a sign that changes from $+1$
on the right side of the product (corresponding to the r.h.s. of (\ref{contactWCF})) to $-1$
on the left side (corresponding to the inverse of the l.h.s. of (\ref{contactWCF})).
The total shift of $\tilde\alpha$ so obtained can be written as
\be
\Delta\tilde\alpha= \frac{1}{2\pi^2} \sum_s \epsilon_s \Omega(\gamma_s)
L_{\qrDg{s}}\left(\cX_{\gamma_s}(s)\right),
\ee
where $\cX_{\gamma_s}(s)=U_{\gamma_{s-1}} \circ U_{\gamma_{s-2}} \circ \cdots \circ U_{\gamma_1}
\cdot \cX_{\gamma_s}$.
In \cite{Alexandrov:2011ac} it was  shown that the quantization property $\Delta\tilde\alpha\in 2\IZ$
follows from the motivic wall-crossing formula of Kontsevich and Soibelman \cite{ks}
in an appropriate classical limit.
This formula generalizes various known and conjectural identities for the Rogers dilogarithm
associated with cluster algebras of Dynkin quivers
(see, e.g., \cite{Gliozzi:1994cs,MR2804544}).

Let us end this discussion with an important remark.
Although the above construction formally gives a satisfactory solution to the wall-crossing problem
in the hypermultiplet sector of type IIA string theory on a CY threefold $X$,
it ignores a crucial problem, namely the  exponential growth $\Omega(\gamma)\sim e^{S(\gamma)}$ of the DT invariants
for large charges, where $S(\gamma)$ is the entropy of a 4D BPS black hole of charge $\gamma$.
Since $S(\gamma)$ scales quadratically in $\gamma$ for large classical black holes, any
generating function of the form $\sum_\gamma \Omega(\gamma) q^{\gamma} $
is divergent, hence making the integral equation (\ref{TBA}) ill-defined.
It was observed in \cite{Pioline:2009ia} that the ambiguity in such divergent sums is of
the same order $e^{-1/g_s^2}$ as NS5-instanton effects, which might therefore cure this problem.
We return to NS5-instantons in \S\ref{sec_NS5}, after discussing S-duality in the
presence of D3-instantons.

\section{S-duality, D3-instantons and mock theta series \label{sec_D3}}

The invariance of  ten-dimensional type IIB string theory under S-duality is a well supported fact
(see e.g. \cite{Green:1997tv} and much subsequent work). It is important to test whether
it continues to hold in vacua with less supersymmetry, in particular in type IIB compactified
on a CY threefold $\hat X$. Assuming that it does, one may e.g. deduce D1-D(-1)
instanton corrections from worldsheet instantons \cite{RoblesLlana:2006is, Alexandrov:2009qq},
obtaining  the first hint
of the form of D-instanton corrections to the HM metric,
or NS5-instantons from D5 \cite{Alexandrov:2010ca}, as we discuss in \S\ref{sec_NS5}.
Here, we focus on the intermediate case of
D3-instantons, which are singlets of $SL(2,\mathbb{Z})$ and should therefore preserve
S-duality by themselves \cite{Alexandrov:2012bu,Alexandrov:2012au}. We shall
show that this is indeed the case thanks to special modular properties of D3-D1-D(-1)
Donaldson-Thomas invariants and of certain indefinite theta series.

\subsection{S-duality in twistor space}
\label{subsec-Stw}

We start by discussing general constraints imposed by S-duality on the twistor space construction.
By a suitable choice of coordinates, we can assume that, even after the inclusion of instanton corrections,
$SL(2,\mathbb{Z})$ acts on $\hat\cM_H(\hat{X})$  by \eqref{SL2Z}. This action must lift to
a holomorphic action on the twistor space $\cZ$. At the classical level, using the Darboux
coordinates \eqref{Darbouxpert} with $F=F^{\rm cl}$ \eqref{classprep},
$r = \tfrac{\tau_2^2}{16}\, e^{-\cK} $ and $c=0$, one can check that \eqref{SL2Z} lifts to
the complex contact transformation \cite{Alexandrov:2008gh}
\be
\label{SL2Zxi}
\begin{split}
&
\xi^0 \mapsto \frac{a \xi^0 +b}{c \xi^0 + d} ,
\qquad
\xi^a \mapsto \frac{\xi^a}{c\xi^0+d} ,
\qquad
\txi'_a \mapsto \txi'_a +  \frac{ c}{2(c \xi^0+d)} \kappa_{abc} \xi^b \xi^c-c_{2,a}\,\varepsilon(g),
\\
&
\begin{pmatrix} \txi'_0 \\ \alpha' \end{pmatrix} \mapsto
\begin{pmatrix} d & -c \\ -b & a  \end{pmatrix} \cdot
\begin{pmatrix} \txi'_0 \\  \alpha' \end{pmatrix}
+ \frac{1}{6}\, \kappa_{abc} \xi^a\xi^b\xi^c
\begin{pmatrix}
c^2/(c \xi^0+d)\\
-[ c^2 (a\xi^0 + b)+2 c] / (c \xi^0+d)^2
\end{pmatrix} ,
\end{split}
\ee
provided the action  \eqref{SL2Z} on $\cM_H$ is supplemented by a suitable action on the $\CP$ fiber, e.g. in the gauge $X^0=1$,
\be
\label{zSL2}
z\mapsto \frac{c\bar\tau+d}{|c\tau+d|}\, z,
\qquad
z\equiv \frac{t+\I}{t-\I}.
\ee

Beyond the classical limit, the Darboux coordinates are no longer given by the simple formulae
\eqref{Darbouxpert}, however, they should still transform as in \eqref{SL2Zxi} up to local contact transformations,
if S-duality is to remain unbroken. This in turn constrains the transformations of the transition functions
on overlapping patches. The constraint can be easily formulated by considering
a covering by an infinite set of open patches $\cU_{m,n}$, which are mapped to each other by S-duality,
including a S-duality invariant patch  $\cU_0\equiv \cU_{0,0}$ \cite{Alexandrov:2009qq,Alexandrov:2012bu}.
Then at the linearized level S-duality
requires that the generating functions $H_{m,n}$ of the contact transformations from $\cU_{0}$
to $\cU_{m,n}$ transform as
\be
H_{m,n}\ \mapsto \frac{H_{m',n'}}{c\xi^0+d}+{\rm reg.},
\qquad  \( m'\atop n'\) =
\(
\begin{array}{cc}
a & c
\\
b & d
\end{array}
\)
\( m \atop n \).
\label{condSH}
\ee
Heuristically, ignoring the fact that the functions $H_{m,n}$ are attached to different (pairs of) patches,
the constraint \eqref{condSH} says that  the formal sum  $\sum H_{m,n}$ should transform as a holomorphic modular form
of weight $-1$.
 In \cite{Alexandrov:2012bu} this constraint was promoted to the full non-linear level
under the assumption that the QK space has two commuting continuous isometries,
which holds for the D3-instanton corrected HM moduli space having two continuous isometries
along fivebrane axions. How to drop this assumption and describe an arbitrary QK space
carrying an isometric action of $SL(2,\mathbb{Z})$ was understood in \cite{Alexandrov:2013mha,Alexandrov:2014mfa}.

\subsection{Modularity of DT invariants}
\label{subsec:modularity}

Before discussing the S-duality invariance of the  D3-instanton corrected metric, let us first
recall the modular properties of DT invariants
associated to dimension 2 sheaves. The same invariants control
D4-D2-D0 and M5-brane black holes, which have been the subject of much research
\cite{Maldacena:1997de,Gaiotto:2006ns,Gaiotto:2006wm,deBoer:2006vg,Denef:2007vg,Dabholkar:2005dt}.
As discussed in \S\ref{stabDT}, given a coherent sheaf $\mathscr{E}$ on $\hat X$,
 D-brane charges are
components of the generalized Mukai charge vector $\gamma$:
\be
\gamma= \ch (\mathscr{E}) \, \sqrt{\Td \hat X}
= p^0 + p^a \omega_a - q'_a \omega^a + q_0'\, \omega_{\hat X}\, ,
\ee
where $\{\omega_a\}$, $\{\omega^a\}$ and $\omega_{\hat X}$ are
respectively a basis of 2-forms, 4-forms and the volume form of $\hat X$, and
$q'_{\Lambda}=q_\Lambda-A_{\Lambda\Sigma}p^\Sigma$ are in general
non-integral  \cite{Alexandrov:2010ca,Alexandrov:2012au}. We shall denote the corresponding
DT invariants by $\Omega(p^0,p^a,q'_a,q'_0;z)$. Dimension-one or zero sheaves ($p^a=p^0=0$)
correspond to D1-D(-1) instantons. As shown in \cite{RoblesLlana:2006is},
the D1-D(-1)-instanton corrected metric is invariant under $SL(2,\IZ)$ provided
\be
\Omega(0,0,0,q_0) = -\chi(\hat X)\, ,
\qquad
\Omega(0,0,q_a,q_0) = n^{(0)}_{q_a}\, ,
\ee
where $n^{(0)}_{q_a}$ are the genus 0 Gopakumar-Vafa invariants governing
the worldsheet instanton corrections. D3-brane instantons correspond to
dimension-two sheaves ($p^0=0,p^a\neq 0$), supported on a divisor $\cD\subset \hat X$.
Dimension-three sheaves will be discussed in \S\ref{sec_NS5_D5}.

We assume that $\cD$ is an ample divisor, i.e. that $[\cD]$ belongs to the \kahler cone.
The intersection matrix of 2-cycles of an ample
divisor provides a natural quadratic form $\kappa_{abc} p^c$ on
$\Lambda=H_4(\hat X,\mathbb{Z})$, with signature $(1,
b_2(\hat X)-1)$. This also provides a quadratic form
$\kappa^{ab}=(\kappa_{abc}p^c)^{-1}$ on $\Lambda^*$.
In the following, we shall use this quadratic form
to identify $\Lambda$ as a sublattice of $\Lambda^*$,
$k^a \mapsto k_a\equiv \kappa_{abc} p^b k^c$.
In the following we shall denote the vector
$(k^1,\dots,k^{b_2(\hat X)})$ as  $\bfk$. For a general
$\bfk\in\Lambda$, the vectors $\bfk_\pm\in \Lambda\otimes \mathbb{R}$ are projections of $\bfk$ onto the positive
and negative definite subspaces of $\Lambda\otimes \mathbb{R}$ defined by the magnetic
charge vector $\bfp$ and the \kahler moduli $\bft$:
\be
\label{defkpm1}
\bfk_+=\frac{\bfk\cdot \bft }{\bfp \cdot \bft^2}\, \bft,
\qquad
\bfk_-=\bfk-\bfk_+,
\qquad
\bfk^2 = \bfk_+^2+\bfk_-^2,
\ee
which satisfy $\bfk_+^2>0$, $\bfk_-^2<0$ for all $\bfk \neq \bf 0$.
We also use the notation $k_+$ to denote the modulus of the vector $\bfk_+$.

As explained in \S\ref{stabDT}, the DT invariants $\Omega(\gamma;\bfz)$ are
piecewise constant in K\"ahler moduli, but can be discontinuous across
walls of marginal stability. This moduli dependence persists in the
large volume limit, and
complicates the analysis of the modular properties of $\Omega(\gamma;\bfz)$
\cite{Manschot:2009ia, Manschot:2010qz}. To deal with this problem, we
express the DT invariants $\Omega(\gamma;\bfz)$ in terms of the `MSW
invariants' $\Omega_\bfp(\bfq',q'_0)$. The latter coincide with the DT invariants
at the so-called `large volume attractor point'
\cite{deBoer:2008fk}\footnote{Note that for CY threefolds
with $b_2(\hat X)=1$, the walls of marginal stability for D3-instantons do not extend
to large volume regime, hence  the MSW and
DT invariants coincide. }
\be
\Omega_\bfp(\bfq',q'_0) = \Omega\(0,\bfp,\bfq',q'_0; \bfz_{\infty}(\gamma) \),
\qquad
\bfz_{\infty}(\gamma)=
\lim_{\lambda\to +\infty}\(\bfb_{}(\gamma)+\I\lambda \, \bft_{}(\gamma)\),
\ee
where $\bfz(\gamma)=\bfb(\gamma)+\I \bft(\gamma)$ is the
standard attractor point. Away from the large volume attractor point $\bfz_{\infty}(\gamma)$ (but still at large
volume), the DT invariant $\Omega(0,\bfp,\bfq',q'_0; \bfz)$ differs from the MSW
invariant $\Omega_{\bfp}(\bfq', q_0')$ by terms of higher order in the MSW invariants
 \cite{Manschot:2009ia, Manschot:2010xp}. The higher order terms
can be thought of as describing bound states of the MSW constituents,
which exist away from the large volume attractor point $\bfz_{\infty}(\gamma)$.
This decomposition is analogous to the decomposition of
the index in terms of the multi-centered black hole bound states.
As shown in \cite{Manschot:2009ia}, the `two-centered' contribution
leads to a modular invariant partition function with the same modular properties as the
elliptic genus, and it is expected that modularity persists to all
orders in the MSW invariants. We stress
that the expansion in MSW invariants is not a Taylor expansion in a
small parameter, rather it is a  finite sum in any chamber separated
 by a finite number of walls from the large volume attractor chamber.
In the remainder of this work, we shall be concerned with only the first term in this expansion, which we
call the `one-instanton approximation' or 'dilute instanton approximation'.

As the name suggests, the MSW invariants are the BPS indices of
the MSW $(0,4)$ superconformal field theory (SCFT) describing D4-brane or M5-branes
wrapped on the divisor $\cD$ \cite{Maldacena:1997de}.
They are unchanged by the `spectral flow' transformations of the charges:
\be
\label{flow}
\bfp \mapsto \bfp ,
\qquad
\bfq'\mapsto \bfq' - \bfeps ,
\qquad
q'_0\mapsto q'_0- \bfeps\cdot \bfq' + \frac12\, \bfp \cdot \bfeps^2,
\ee
which are induced by monodromies around the large volume point of $\cM_H(\hat
X)$, $\bfz\to \bfz+\bfeps$, and leave
\be
\label{defqhat0}
\hat q_0 \equiv
q_0' -\frac12\, \bfq'^2
\ee
invariant. Decomposing
\be
\bfq' = \bfmu + \bfeps +\tfrac12\,\bfp ,
\ee
where  $\bfmu \in \Lambda^*/\Lambda$ is the residue
class of $\bfq'-\tfrac12\bfp$ modulo $ \bfeps$, it follows that the MSW invariant
$\Omega_\bfp(\bfq',q'_0)\equiv \Omega_{\bfp,\bfmu}(\hat q_0)$
depends only on $\bfp$, $\bfmu$ and $\hat q_0$.
The partition function of MSW invariants for fixed divisor $\cD$ is
the elliptic genus of the SCFT,
\begin{eqnarray}
\label{eq:ellgenus}
\mathcal{Z}_{\bfp}(\tau,\bfy) &=& \sum_{\bfq', q'_0} \, (-1)^{\bfp\cdot \bfq}\, \bOm_\bfp(\bfq',q'_0)\,
\expe{-\hat q_0 \tau  - {1\over 2} \,\bfq'^2_- \tau - {1\over 2}\,\bfq'^2_+ \bar \tau +\bfq'\cdot \bfy}
\end{eqnarray}
with $\bfy\in \Lambda\otimes \mathbb{C}$ and $\bOm_\bfp(\bfq',q'_0)$
the rational MSW invariant defined analogously to \eqref{ratDT}.
When $\gamma$ is primitive, it follows from general properties
of the MSW SCFT that the elliptic genus  \eqref{eq:ellgenus} is a
multi-variable Jacobi form of weight $(-\tfrac32,\frac12)$ under $SL(2,\IZ)$,
with multiplier system $M_{\cZ}=\expe{\varepsilon(\trans)\,{\bfch}\cdot \bfp}$,
where $\varepsilon(\trans)$ is as in \eqref{multeta}. If
$\bfp$ is not a primitive vector, the BPS indices might not be
related to a proper CFT due to states at threshold stability. However,
wall-crossing  arguments \cite{Manschot:2010qz} and explicit
calculations \cite{Manschot:2011dj} suggest that nevertheless the generating function of
$\bOm_\bfp(\bfq',q'_0)$ exhibits good modular properties under
$SL(2,\mathbb{Z})$.

The invariance of $\bOm_\bfp(\bfq',q'_0)$ under the spectral flow \eqref{flow}
implies that the elliptic genus has a theta function decomposition:
\be
\label{thetaeg}
\mathcal{Z}_\bfp (\tau,\bfy,\bft) = \sum_{\bfmu \in \Lambda^*/\Lambda}
h_{\bfp,\bfmu} (\tau)\, \overline{\theta_{\bfp,\bfmu} (\tau,\bfy,\bft)}\, ,
\ee
where $\theta_{\bfp,\bfmu}$
is the Siegel-Narain theta series
 associated to the lattice $\Lambda$ equipped
with the  quadratic form $\kappa_{ab}$ of signature $(1,b_2(\hat X)-1)$,
\be
\label{defth}
\theta_{\bfp,\bfmu} (\tau,\bfy,\bft) =
\sum_{\bfk\in \Lambda+\bfmu+ \tfrac12 \bfp}
 \signkp\,\expe{ \frac12\, \bfk_+^2 \tau+\frac12\, \bfk_-^2 \bar\tau
+ \bfk\cdot \bfy},
\ee
and the $\bfy$-independent coefficients are given by
\be
\label{defchimu}
h_{\bfp,\bfmu}(\tau) = \sum_{\hat q_0 \leq r \chi(\cD)/24}
\bOm_{\bfp,\bfmu}(\hat q_0)\,
\expe{-\hat q_0 \tau }.
\ee
Since the theta series \eqref{defth} is a vector valued Jacobi form of
modular weight $(\tfrac12,\tfrac{b_2(\hat X)-1}{2})$ and multiplier system
$M_\theta$ under $SL(2,\IZ)$, it follows that $h_{\bfp,\bfmu}$
must transform as a vector-valued
holomorphic\footnote{\label{foot:mock}Non-compact directions in the target space of the
CFT could potentially lead to mock modular forms and thus holomorphic anomalies \cite{Troost:2010ud}. For
local CY manifolds, e.g. $\cO(-K_{\mathbb{P}^2})\to
\mathbb{P}^2$, it is known that the holomorphic generating series of DT-invariants
for sheaves with rank $>1$ requires a non-holomorphic addition in order to transform as a
modular form \cite{Vafa:1994tf}. We assume that this issue does not arise here.}
modular form of negative weight $(-\frac{b_2(\hat X)}{2}-1,0)$
and  multiplier system $M(\trans)=M_{\cZ}\times \overline{M_\theta}^{-1}$
under the full modular group $SL(2,\IZ)$. The latter is equivalent to
$M(\trans)=M_{\cZ}\times M_\theta$ since $M_\theta$ is unitary.

\subsection{Dilute D3-instantons}
\label{D3instanton}

Let us now explain why D3-D1-D(-1)   instanton corrections
are consistent with S-duality, at least in the dilute instanton approximation.
To this end it is sufficient to consider the contribution of a single
homology class $\bfp$, but to include the sum over embedded classes
$\bfq$ and $q_0$. This avoids the complications arising from the divergent sum over $\bfp\in
H^4(\hat X,\mathbb{Z})$, as mentioned at the end of \S \ref{sec_DinstTwistor}.

\subsubsection{Transition functions}

First, we observe that the twistor description of D3-instantons satisfies the requirement of \S\ref{subsec-Stw}, namely
that the formal sum of  transition functions  transforms as a holomorphic
modular form of weight $-1$.
The transition functions  generating  the contact transformations \eqref{KStrans2}
are
\be
H_\gamma=\frac{\Omega(\gamma)}{(2\pi)^2} \,
\Li_2(\qrD \cX_\gamma)+\cdots,
\label{D3transfun}
\ee
where $\gamma=(0,\bfp,\bfq,q_0)$ and
$\cdots$ denote terms non-linear in the DT invariants whose explicit form can be found in \cite{Alexandrov:2009zh}.
In the leading term one can replace $\Li_2(x)$ by $x$ at the cost
of replacing
the integer DT invariants by their rational counterparts \eqref{ratDT}. Furthermore, in the dilute instanton approximation
the latter can be replaced by the MSW invariants $\bOm_\bfp(\bfq',q'_0)$ introduced
in the previous subsection. Choosing $\qrD=(-1)^{\bfp\cdot \bfq}$, we arrive at the formal
sum
\be
\label{eqHp}
H_\bfp=\sum_{\bfq', q'_0}\bH_\gamma,
\qquad
\bH_\gamma= \frac{(-1)^{\bfp\cdot \bfq'}}{(2\pi)^2}\,
\bOm_{\bfp,\bfmu}(\hat q_0)\,
\expe{ \bfp\cdot\bftxip - \bfq' \cdot \bfxi - q'_0 \xi^0} .
\ee
Similarly to \eqref{thetaeg}, using monodromy invariance the sum can be rewritten as
\be
\label{Hdec}
H_\bfp= \frac{1}{(2\pi)^2}\,
\expe{ \bfp\cdot \bftxip }
\sum_{\bfmu\in \Lambda^*/\Lambda}
h_{\bfp,\bfmu} (\xi^0)\, \ZTheta_{\bfp,\bfmu} (\xi^0,\bfxi)\, ,
\ee
where
$h_{\bfp,\bfmu}(\xi^0)$ is the modular function defined in \eqref{defchimu},
now evaluated at $\tau=\xi^0$, and $\ZTheta_{\bfp,\bfmu} (\xi^0,\bfxi)$
is a holomorphic theta series defined by the quadratic form $-\kappa_{ab}$
which transforms formally as a holomorphic Jacobi form of weight
$\tfrac{b_2}{2}$, multiplier system $M_\theta^{-1}$ and index $m_{ab}=-\frac12 \kappa_{ab}$ \cite{Alexandrov:2012au}.
Finally, thanks to the term proportional to $c_{2,a}$ in the transformation of $\bftxip$,
the exponential prefactor in \eqref{Hdec} transforms as the automorphy
factor of a multi-variable holomorphic Jacobi theta series with the index
$m_{ab}=\frac12 \kappa_{ab}$ and multiplier system $M_{\cZ}^{-1}$.
Combining the transformation properties of all factors, we conclude that,
under the action \eqref{SL2Zxi}, the formal sum \eqref{eqHp} indeed transforms as a
holomorphic Jacobi form of weight $-1$ and trivial multiplier system, as required by S-duality.

However, this analysis overlooks the important fact that the quadratic form $\kappa_{ab}$ has
indefinite signature $(1,b_2-1)$, and therefore the theta series $\ZTheta_{\bfp,\bfmu} (\xi^0,\bfxi)$
is  divergent. Fortunately, it never actually arises
as such in the computation of the metric, rather each of the terms in \eqref{eqHp} must
be integrated along a different contour, which renders the resulting series convergent.
To put the above heuristic argument on solid ground, one should analyze the transformation
properties of the quantities which enter the computation of the metric,
namely the Darboux coordinates and the contact potential. Below we restrict to the contact potential
and Darboux coordinate $\bfxi$, which exhibit the key mechanism, referring the reader
to  \cite{Alexandrov:2012au} for a complete discussion.

\subsubsection{Contact potential}

The contact potential $e^{\phi}$ is given in \eqref{contactpot-D}.
The first term on the right hand side is the classical term, with
$\frac{1}{8}e^{-\cK(z,\bz)}$ equal to the volume $\cV={1\over 6}\bft^3$ of $\hat X$. Using the transformation
properties of $\tau_2$ and $\bft$, one easily checks that the
classical part transforms with weight $(-\half, -\half)$.
The instanton corrections involve the Darboux coordinates
$\cX_\gamma$, which are solutions to the integral equations
(\ref{TBA}). In order to relate these corrections to the modular functions
of the previous section, we start by writing the 3rd term on the right hand side of
\eqref{contactpot-D} as:
\be
\label{eq:instH}
\delta_\bfp \,e^{\phi}=-\frac{\I\tau_2}{16}
\sum_{q_\Lambda}
\int_{\ellg{\gamma}}
\frac{\d \varpi}{\varpi}\,\( \varpi^{-1} Z_\gamma -\varpi\bZ_\gamma\)
\bH_\gamma(\xi^0,\bfxi,\tilde \bfxi),
\ee
where $\bH_\gamma$ is given in \eqref{eqHp}.
In the dilute instanton approximation we can replace the Darboux coordinates appearing in the arguments
of $\bH_\gamma$ by their classical expressions \eqref{Darbouxpert}
with the prepotential \eqref{classprep}. Furthermore, we keep only the leading terms in
the limit where $\bft\to\infty$ and  the product $z\, \bft$ remains constant (where $z$
is the Cayley-rotated coordinate on $\CP$, see \eqref{zSL2}).
This is motivated by the fact that the saddle points of the integral lie at
$z_\gamma=-\I (k+b)_+/\sqrt{\bfp\cdot \bft^2}$ for
$\bfp\cdot\bft^2>0$.  As a result,  $\bH_\gamma(\xi^0,\bfxi,\tilde \bfxi)$ simplifies to the following form
 \cite{Alexandrov:2012au}:
\be
\begin{split}
\bH_{\gammap} =& \frac{(-1)^{\bfp\cdot\bfq}}{(2\pi)^2}\, \bOm_{\bfp,\bfmu}( \hat q_0) \,
\expe{\I S_{\rm cl} -\frac12  (\bfq+ \bfb)_+^2 \,
\bar\tau - \(\hat q_0+ \frac12 (\bfq+ \bfb)_-^2\)\tau+\bfc \cdot(\bfq +\frac12 \bfb)
+\I Q_{\gammap}(z)},
\end{split}
\ee
where $S_{\rm cl}$ is the leading part of the Euclidean D3-instanton action
in the large volume limit, and  $Q_{\gamma}(z)$ is the only part which depends on
the fiber coordinate $z$,
\be
S_{\rm cl} = \frac{\tau_2}{2}\, \bfp\cdot \bft ^2 - \I\, \bfcl\cdot \bfp ,
\qquad
Q_{\gammap}(z)=\tau_2 \, \bfp\cdot \bft ^2\,
\left(z +\I\, \frac{ \kbp}{\sqrt{\bfp\cdot \bft ^2}}\right)^2.
\label{clactinst}
\ee
Keeping the leading contributions to the remaining terms in
(\ref{eq:instH}) leads to
\be
\delta_{\bfp}\, e^{\phi}=-\frac{\tau_2}{4} \sum_{q_\Lambda}
\int_{\ell_\gamma} dz \left[\hat q_0+{1\over 2} (\bfk+\bfb-\I z \bft)\cdot (\bfk +\bfb -3\I z \bft) \right] H_{\gamma}
+ {\rm c.c.}
\label{resPhi0}
\ee
The  integral over $z$ is now Gaussian, leading to
\be
\begin{split}
\delta_{\bfp}\, e^\phi = &
\frac{\tau_2\, e^{-2\pi S_{\rm cl}}}{16\pi^2\sqrt{2\tau_2\, \bfp\cdot \bft^2}}\,
\cD_{-\frac{3}{2}}\, \sum_{\bfmu\in\Lambda^*/\Lambda}
h_{\bfp,\bfmu}(\tau)\,\overline{\theta_{\bfp,\bfmu}(\tau,\bft,\bfb,\bfc)} +{\rm c.c.},
\end{split}
\label{resPhi}
\ee
where $\theta_{\bfp,\bfmu}(\tau,\bft,\bfb,\bfc)$
is a  theta function similar to \eqref{defth}, which transforms as
a vector-valued modular
form with weight $\half (1,b_2(\hat X)-1)$ and  multiplier
system $M_\theta$, see \cite[Eq. (A.6)]{Alexandrov:2012au}.
The action of $\cD_{-\frac{3}{2}}={1 \over 2\pi \I} \left(\partial_\tau-\frac{3/2}{2\I\tau_2} \right)$
raises the modular weight from $( -\tfrac32,\half)$ to $(\half,\half)$, while the overall factor of
$\tau_2$ reduces this to $(-\half,-\half)$. The transformation
property of $\tilde c_a$ \eqref{SL2Z} is now seen to cancel
the non-trivial phases due to the multiplier system
$M_\mathcal{Z}$ of $\mathcal{Z}_\bfp(\tau,\bfy,\bft)$, establishing that  the instanton correction
$\delta_{\bfp}\, e^{\phi}$ transforms correctly under S-duality.

\subsubsection{Darboux coordinates and Eichler integrals}

Our next example is the coordinate $\bfxi$ on twistor space, which
has the general form:
\be
{\bfxi}={\boldsymbol \zeta}+{\tau_2 \over 2}\left( { \bfz \over t}-\bar \bfz t\right)
+\frac{1}{8 \pi^2}\sum\limits_{\gamma} \Omega(\gamma;\bfz)\, \bfp
\int_{\ellg{\gamma}}\frac{\d \varpi'}{\varpi'} \frac{\varpi+\varpi'}{\varpi-\varpi'}\,
\log\left[1-\sigma_\gamma \cX_\gamma  \right].
\ee
As explained in \cite{Alexandrov:2012bu}, the S-duality transformations of such expressions
can be greatly simplified by modifying the integration kernel by a $t$-independent term, at the cost of
correcting the mirror map \eqref{symptobd}. After this change and in the large volume limit,
$\delta_{\bfp}\, \bfxi$ becomes
\be
\delta_{\bfp}\, \bfxi=\bfp\sum_{q_\Lambda} \int_{\ell_\gamma}
\frac{dz'}{z'-z} \,\bH_\gamma .
\ee
Unlike \eqref{resPhi0}, the integral is no longer Gaussian but can
be expressed as an Eichler integral
\be
\label{xiazperiod}
\delta_{\bfp}\,\bfxi=-\frac{\, e^{-2\pi S_{\rm cl}}}{4\pi}\bfp
\sum_{\bfmu\in\Lambda^*/\Lambda}
h_{\bfp,\bfmu}(\tau)\,
\int_{\bar\tau}^{-\I \infty}
\frac{\overline{\BTheta_{\bfmu} (w,\bar\tau;\bz)}\, \de \bar w}{\sqrt{\I(\bar w-\tau)}} ,
\ee
where $\BTheta_{\bfmu} (w,\bar\tau;\bz)$ (given \cite[Eq. (4.24)]{Alexandrov:2012au})
is a modular form of weight $({3 \over 2}, {b_2-1 \over 2})$. Such
integrals transform inhomogeneously under $SL(2,\mathbb{Z})$,
\be
\label{modcJ}
\delta_\bfp\, \bfxi \mapsto (c\tau+d)^{-1} \left(
\delta_\bfp\, \bfxi  +\frac{\,e^{-2\pi S_{\rm cl}}}{4\pi} \bfp\sum_{\bfmu\in\Lambda^*/\Lambda}
h_{\bfp,\bfmu}(\tau)\,
\int_{-d/c}^{\I\infty} \frac{\overline{\BTheta_{\bfmu} (w,\bar\tau;\bz)}\,
\de \bar w }{[\I(\bar w-\tau)]^{1/2}}\right) .
\ee
The overall weight $(-1,0)$ is in agreement with the classical
transformation of $\bfxi$ (\ref{SL2Zxi}), however the period integral $\int_{-d/c}^{\I\infty} \de\bar w$
makes this transformation anomalous. Remarkably, using standard techniques for mock theta series
\cite{Zwegers-thesis} one can show that
\be
\delta_\bfp\, \widehat{\bfxi}=\delta_\bfp\, \bfxi -2\pi \I \,\bfp \, H_{\rm anom}
\ee
does transform as a modular form of weight $(-1,0)$. Here $H_{\rm
  anom}$ is an indefinite theta function:
\be
H_{\rm anom}=\half \sum_{q_\Lambda} \left[ \sgn\((\bfk+\bfb)\cdot \bft\) - \sgn\((\bfk+\bfb)\cdot \bft'\)\right] \bH_\gamma,
\ee
with $\bft'$ lying on the boundary of the K\"ahler cone.
$H_{\rm anom}$ transforms also by a period integral, precisely cancelling the
one of $\delta_\bfp\, {\bfxi}$. More generally, $H_{\rm anom}$ generates a contact
transformation which precisely cancels the modular anomaly in all Darboux coordinates,
thus establishing the S-duality invariance of the D3-instanton corrected HM metric.
It would be very interesting to derive $H_{\rm anom}$
and the choice of $\bft'$  from  first principles.

\section{Toward NS5-instanton effects\label{sec_NS5}}

In addition to D-instanton effects, the metric on the HM moduli space $\cM_H$
receives instanton corrections of  order $e^{-1/g_s^2}$ from Euclidean NS5-branes
wrapping the whole threefold $X$ \cite{Becker:1995kb}. On the type IIB side, it is clear
that such corrections are  necessary to restore S-duality, since D5 and NS5 branes transform as a doublet
under $SL(2,\IZ)$.  For $k$ NS5-branes the  expected correction is, schematically,
\be
g_{{}_{\rm NS5}}\sim e^{-4\pi|k|r-\pi i k \sigma} \mathscr{Z}_{k},
\label{NS5corr}
\ee
where $\mathscr{Z}_k$ is the partition function for the degrees of
freedom localized on the NS5-branes and $r\sim  \cV/g_s^2$. In particular, unlike D-instanton effects,
these corrections break the continuous translational symmetry of the NS-axion to a discrete subgroup
$\sigma \mapsto \sigma +2\kappa$, $\kappa\in \mathbb{Z}$, as anticipated in equation (\ref{heis0}).
Our goal in this section is to infer the form of
$\mathscr{Z}_{k}$ (more precisely, the corresponding correction to the twistor space $\cZ$)
from the known D5-instanton corrections, at least in a linearized approximation.
Before doing this, we discuss  constraints on $\mathscr{Z}_{k}$ coming from the topology
of the axion circle bundle $\mathscr{C}_{\sigma}$. We close with some speculative comments
on relations between NS5-instantons and quantum integrable systems.

\subsection{Topology of the axion circle bundle}
\label{sec-topol}

\subsubsection{NS5-partition function and theta series}

The fivebrane  partition function $\mathscr{Z}_{k}$ is in general a function of the dilaton $\phi$,
complex structure  moduli $z^a$ and RR moduli $C$ (in  type IIA variables). Its dependence
on $C$ is  strongly constrained by the fact that the corrected
metric must stay invariant under the large gauge transformations \eqref{heis0}.
In view of \eqref{NS5corr}, this implies that, under an integer shift $C\mapsto C+H$ with $H\in H^3(X, \mathbb{Z})$,
\be
\mathscr{Z}_{k}(C+H)= \big(\lambda(H)\big)^k\,
\expe{\tfrac{k}{2}\left<C, H\right>} \mathscr{Z}_{k}(C).
\label{NS5periodicity}
\ee
In  words, $\mathscr{Z}_{k}$ must be a section of the theta line bundle
$(\mathscr{L}_\Theta)^{\otimes k}$ over $\cT$, with first Chern class
$c_1(\mathscr{L}_\Theta)=\omega_\cT$ (see \eqref{Chcl}). In the Weil complex structure,
 $(\mathscr{L}_\Theta)^{\otimes k}$ is known to admit $|k|^{b_3(X)/2}$ holomorphic sections
\cite{Witten:1996hc}, corresponding to the Siegel theta series
\be
\vartheta_{k, \mu}(C)=\sum_{n\in \Gamma_m+\mu+\theta}
\expe{\tfrac{k}{2}(\zeta^{\Lambda}-n^{\Lambda})\bar{\cN}_{\Lambda\Sigma}
(\zeta^{\Sigma}-n^{\Sigma})+k(\tilde\zeta_\Lambda-\phi_\Lambda)n^{\Lambda}+\tfrac{k}{2}
(\theta^{\Lambda}\phi_\Lambda-\zeta^{\Lambda}\tilde\zeta_\Lambda)}
\label{GaussianNS5}
\ee
labelled by vectors $\mu\in \Gamma_m/|k|\Gamma_m$, where $\Gamma_m$
is a Lagrangian sublattice of $H^3(X, \mathbb{Z})$, here spanned by A-cycles.
Physically, the sum over $n^\Lambda$ labels the topological sectors of the (imaginary)
self-dual 3-form field strength $\cH$ living on $k$  fivebranes. Indeed,
the Siegel theta series can be obtained by holomorphic factorization
of the (non-holomorphic) partition function $\mathscr{Z}_{k, 3-\mathrm{form}}$
of a Gaussian 3-form on $X$
\cite{Witten:1996hc,Henningson:1999dm,Dijkgraaf:2002ac,Moore:2004jv,Belov:2006jd}:
$$
\mathscr{Z}_{k, 3-\mathrm{form}}\sim \sum_{\mu\in \Gamma_m/|k|\Gamma_m} \vartheta_{k, \mu}(C)\, \overline{\vartheta_{k, \mu}(C)}.
$$

In general however, the chiral fivebrane  worldvolume  theory is non-Gaussian, and the only conclusion that
can be drawn from \eqref{NS5periodicity} is that $\mathscr{Z}_{k}$ must be a linear combination
of non-Gaussian theta series
\be
\mathscr{Z}_{k}(C) =\sum_{\mu\in \Gamma_m/(|k|\Gamma_m)}
\sum_{n\in \Gamma_m+\mu+\theta} \Psi_{k,\mu}
( \zeta^{\Lambda}-n^{\Lambda}) \expe{k(\tilde\zeta_\Lambda-\phi_\Lambda)
n^{\Lambda}+\tfrac{k}{2}(\theta^{\Lambda}\phi_\Lambda-\zeta^{\Lambda}\tilde\zeta_\Lambda)},
\label{nonGaussianNS5}
\ee
where we only displayed the dependence on the $C$-field. The $\Psi_{k,\mu}$'s can be interpreted
as wave-functions in the real polarization corresponding to the Lagrangian
subspace of $H_3(X,\IR)$ spanned by periods along the A-cycles.

\subsubsection{Metric dependence}

More generally, for consistency of the correction \eqref{NS5corr}, the partition function $\mathscr{Z}_k$
should be a (not necessarily holomorphic) section of $\mathscr{C}_\sigma^k$,
where $\mathscr{C}_\sigma$ is the circle bundle where the NS axion $e^{\I \pi \sigma}$ is
valued \cite{Alexandrov:2010ca}. As indicated in \eqref{Chcl}, this circle bundle has curvature both over the fiber $\cT$
and base $\cM_C(X)$ of the intermediate Jacobian $\cJ_W(X)$. The curvature over $\cT$
reflects the non-trivial behavior \eqref{heis0}, \eqref{NS5periodicity} under large gauge transformations, while the curvature
over $\cM_C(X)$ shows that under a monodromy $M$ in  $\cM_C(X)$, under which
the holomorphic 3-form transforms as $ \Omega_{3,0} \mapsto e^{f} \Omega_{3,0}$,
the NS-axion and NS5 partition function transform by
\be
\sigma \mapsto \sigma + \frac{\chi(X)}{24\pi} \Im f +2\kappa(M),
\qquad
\mathscr{Z}_k \mapsto e^{\I k \frac{\chi(X)}{24} \Im f + 2\pi i k \kappa(M)} \mathscr{Z}_k,
\label{phases}
\ee
where $\kappa(M)$ is the logarithm of a character of the monodromy group (see \cite{Alexandrov:2010ca} for more details).

The cancellation of phases between $e^{-\I \pi k\sigma}$ and $\mathscr{Z}_k$ is expected
by the general inflow mechanism for local anomalies \cite{Duff:1995wd}.
A complication however is that the first Chern class \eqref{Chcl} is not an integer cohomology class,
rather $2c_1(\mathscr{C}_{\sigma})\in H_2(\cM_H,\IZ)$, since $\tfrac{1}{12}\chi(X)\omega_C$ can be identified as the first Chern class of
the determinant line bundle of the Dirac operator for chiral spinors on $X$ \cite{Witten:1985xe,Monnier:2010ww},
also known as the BCOV line bundle \cite{1167.32016}.
This means that $\sigma$ is only defined up to a half-period, and $\mathscr{Z}_k$ has a sign ambiguity.
This sign ambiguity corresponds to a choice of character $\kappa(M)$ in (\ref{phases}),
and also appears to be related to the 'orientation data' in the theory
of generalized Donaldson-Thomas invariants \cite{ks,2012arXiv1212.3790H}.
We expect that global anomaly cancellation will ensure that these two ambiguities cancel.

\subsection{Fivebrane corrections to the contact structure}
\label{sec_D5NS5}

\subsubsection{NS5-branes from D5-branes}
\label{sec_NS5_D5}

Since (NS5,D5)-branes  transform as a doublet under S-duality, the corrections to the
complex contact structure on $\cZ$ induced by NS5-instantons can in principle be inferred
from the D5-instanton corrections described in \S\ref{sec_DinstTwistor}. This procedure
was implemented at the linearized level in \cite{Alexandrov:2010ca} as follows.
Start from the transition function
\be
\label{HDinstp}
\bH_{\gamma}=\frac{\bOm(\gamma)}{(2\pi)^2}\,\qrD \,
\expe{p^\Lambda\txi'_\Lambda-q'_\Lambda \xi^\Lambda}
\ee
generating the contact transformation across a
BPS ray associated to an instanton with  non-vanishing D5-brane charge $p^0$,
and act on it by an $SL(2,\IZ)$ transformation
$g=(\begin{smallmatrix} a & b \\ c & d  \end{smallmatrix})$,
choosing $(c,d)=(-k/p^0,p/p^0)$ such that $p^0=gcd(k,p)$.
Such a transformation maps a D5-brane into a $(k,p)$-fivebrane, where $k$ labels the NS5-brane charge.
Using \eqref{SL2Zxi} and \eqref{condSH}, one finds that the transition function associated
to such a fivebrane takes the form
\be
\label{5pqZ2}
H_{k,p,\hgamma}=-\frac{\bOm(\gamma)}{(2\pi)^2}\,\frac{k}{p^0}\(\xi^0 -n^0\)\qrD
\expe{k  S_\alpha+ \frac{p^0\( k \hat q_a (\xi^a-n^a) + p^0 \hat q_0\)}{k^2(\xi^0-n^0)}
+ a\,\frac{p^0 q'_0}{k}- c_{2,a} p^a\eps(g)} ,
\ee
where $\hgamma$ denotes the remaining D3-D1-D(-1)-charges and
\be
\label{defSa}
S_\alpha= \alpha + n^\Lambda\txi_\Lambda+
\Fcl(\xi-n)-\hf\,A_{\Lambda\Sigma}n^\Lambda n^\Sigma ,
\qquad (n^0,n^a)=\(\frac{p}{k},\frac{p^a}{k}\)\in \IZ/k .
\ee
This transition function generates a discontinuity in the Darboux coordinates across
the image of the  BPS ray $\ell_\gamma$ under the same $SL(2,\IZ)$ transformation,
namely a meridian  $\ell_{k,p,\hgamma}\subset\CP$ joining the
two roots of the equation $\xi^0(t)=n^0$ where \eqref{5pqZ2} has
essential singularities.

One can perform two consistency checks on the above result.
First, by evaluating the Penrose transform of \eqref{5pqZ2}
\be
\int_{\ell_{k,p,\hgamma}}  \frac{\de\varpi}{\varpi}\,
H_{k,p,\hgamma}(\xi(\varpi),\txi(\varpi),\alpha(\varpi))
\label{Ptrans}
\ee
in the small string coupling limit, which justifies the use of the saddle point approximation, it was checked
\cite{Alexandrov:2010ca} that \eqref{5pqZ2} produces corrections with the correct semi-classical
action known from the analysis of instanton solutions
in $N=2$ supergravity \cite{Looyestijn:2008pg}.
The second check is to verify that  \eqref{5pqZ2} is consistent with large gauge transformations \eqref{heis0}
and monodromy around the large volume point. This requires  that
the set of transition functions should be mapped to itself under the corresponding actions
on twistor space. This is indeed the case,
as shown in \cite{Alexandrov:2010ca} and further clarified in \cite{Alexandrov:2014rca}.
The problem of going beyond the linearized regime and elevating the
infinitesimal contact transformations generated by \eqref{5pqZ2} to their finite counterparts was addressed in 
\cite{Alexandrov:2014mfa,Alexandrov:2014rca}.

\subsubsection{Relation to topological strings}
\label{sec_topstring}

A general prediction of S-duality is that the partition function of a single NS5-brane
($k=1$) on a CY threefold $X$ should be governed by
the (ordinary) DT invariants of $X$ with $p^0=1$, which are in turn related to higher genus
Gromov-Witten invariants \cite{gw-dt}, and therefore to topological strings
\cite{Dijkgraaf:2002ac,Nekrasov:2004js,Kapustin:2004jm}.
A precise relation of this sort follows immediately from
\eqref{5pqZ2}. Indeed, the formal sum over all charges
\be
H_{{\rm NS5}}^{(1)}(\xi,\txi,\alpha)=\sum_{p,p^a, q_\Lambda} H_{1,p,\hgamma}(\xi,\txi,\alpha)
\label{fullH}
\ee
is invariant under Heisenberg transformations \eqref{heis0twistor}.
Thus it can be cast in the form \eqref{nonGaussianNS5} of a non-Gaussian theta series \cite{Alexandrov:2010ca}
\be
\label{hthxin}
H_{{\rm NS5}}^{(1)}(\xi,\txi,\alpha)=\frac{1}{4\pi^2}
\sum_{n^\Lambda }
\cH_{{\rm NS5}}^{(1)}\( \xi^\Lambda - n^\Lambda\)\,
\expe{\alpha+n^\Lambda(\txi_\Lambda-\phi_\Lambda)} .
\ee
Remarkably, the wave-function $\cH_{{\rm NS5}}^{(1)}$ turns to be proportional
(up to $\xi^0$-dependent factors involving the Mac-Mahon function)
to the topological A-model string amplitude on $\hat X$ in the real polarization,
\be
\cH_{{\rm NS5}}^{(1)}(\xi^\Lambda) \sim
 \Psi_\IR^{\rm top}(\xi^\Lambda) .
\label{NS5DTrelation}
\ee
This relation indicates that the proper habitat for the topological string amplitude
is the space $H^1(\cZ_{D},O(2))$ parametrizing deformations of the D-instanton corrected
twistor space.

\subsection{NS5-branes and quantization of cluster varieties}
\label{subsec-cluster}

We conclude our discussion of quantum corrected HM moduli spaces by collecting  several
indications pointing to deep relations with quantum integrable systems. The general relation to
integrable systems is of course built in from the fact that $\cM'_H$, the HK manifold dual to
the perturbative $\cM_H$ by the QK/HK correspondence of \S\ref{sec_QKHK}, is a HK manifold
fibered by algebraic tori, and therefore provides an example of a complex
integrable system.\footnote{See \cite{2013arXiv1303.3253K} for a recent analysis
of the relation between these integrable systems and wall-crossing in Donaldson-Thomas theory.}

More interestingly, this relation seems to extend in the presence of D-instanton corrections,
in view of the fact \cite{Gaiotto:2008cd,Alexandrov:2010pp} that the integral equations \eqref{TBA}
governing the D-instanton corrections to the metric on $\cM_H$ coincide with equations of
Thermodynamic Bethe Ansatz (TBA) typically describing the spectrum of two-dimensional
integrable models \cite{Zamolodchikov:1989cf}. Moreover,
the D-instanton corrected contact potential \eqref{contactpot-D}
is identified with the free energy of the system associated to this TBA,
and the corresponding S-matrix can be shown to satisfy all axioms of integrability \cite{Alexandrov:2010pp}.

The TBA equations can often be rewritten as a system of discrete equations, known as Y-system,
corresponding to gluing conditions \eqref{KStrans2} of our framework.
In particular, the same dilogarithm identities which ensure consistency with
wall-crossing are well-known to arise as a consequence of periodicity of Y-systems
\cite{Ravanini:1992fi}. Their mathematical underpinning
is Fomin and Zelevinsky's theory of \emph{cluster algebras}\footnote{The relation between
wall-crossing and cluster transformations was noted already in \cite{ks},
and further analyzed in the context of $\cN=2$Ê gauge theories \cite{Gaiotto:2009hg}
and $\cN=2$ supergravity \cite{Alexandrov:2011ac}.}
\cite{2001math4151F}, and
its generalization to \emph{cluster varieties}, developed by Fock and Goncharov \cite{2008InMat.175..223F}.
In particular, a structure very similar to NS5-instanton corrections arises
when quantizing \emph{cluster $\cA$-varieties} \cite{FGquant}.
An $N$-dimensional cluster $\cA$-variety is a collection of complex tori $(\mathbb{C}^{\times})^N$
glued together into a symplectic algebraic variety using cluster transformations.
Geometric quantization produces a pre-quantum vector bundle $\mathscr{V}_\hbar \to \cA$,
depending on a rational parameter $\hbar=s/r$, where $s$ is the first Chern class and $r$
encodes the rank of the bundle. For $\hbar=1$, the corresponding line bundle $\mathscr{V}_1$
is described by exactly the same gluing conditions \eqref{KStrans2}
as for the holomorphic bundle $\mathscr{L}_{\cZ'}$ entering the QK/HK-correspondence.
Thus we may view the hyperholomorphic bundle $\mathscr{L}$ as a pre-quantum line
bundle for the geometric quantization of $\cM'$ (as also observed in \cite{Hitchin:2012}).

To explain the relation with NS5-branes, let us for simplicity restrict to the case
of a two-dimensional cluster $\cA$-variety, with local coordinates
$(a,b)\in \mathbb{C}^{\times}\times \mathbb{C}^{\times}$. Sections of $\mathscr{V}_1$
are multivalued functions $F(a,b)$ on $\mathbb{C}^{\times}\times \mathbb{C}^{\times}$,
and by a choice of polarization one may restrict to holomorphic 'wave functions' $\Psi(a)$
on $\mathbb{C}^{\times}$. Fock and Goncharov implement this restriction explicitly via
Fourier expansion in one of the variables \cite{FGquant}:
\be
F(a,b)=\sum_{n\in \mathbb{Z}} \Psi(\log\, a - 2\pi \I n)
\exp \left[-\tfrac{\log\, a \, \log b}{4\pi \I}\right] \, b^{n}.
\label{polarization}
\ee
Identifying $a=e^{2\pi \I \xi}$ and $b=e^{2\pi \I \tilde\xi}$, up to the factor $e^{-\I\pi \tilde \alpha}$,
this takes the same form as the NS5-partition function (\ref{hthxin}).
Thus the twistor space partition function of a single NS5-brane ($k=1$) may be thought
of as a section of a pre-quantum line bundle over an $\cA$-cluster variety.
More generally, for $\hbar \neq 1$, sections of the vector bundle $\mathscr{V}_\hbar$
can be represented by vector-valued functions $\Psi_{\hbar^{-1}, \ell}(\xi)$ via
a generalization of the expansion (\ref{polarization}). In the case $s=1$ and $r=1/\hbar \in \mathbb{Z}_+$,
this becomes
\be
\label{Fhbar}
F_\hbar(\xi, \tilde\xi)=\sum_{\ell\in \mathbb{Z}/(\mathbb{Z}|\hbar|^{-1})}
\sum_{n\in \mathbb{Z}+ \ell |\hbar|} \Psi_{\hbar^{-1}, \ell}(\xi-n) \expe{(2n \txi-\xi\txi)/2\hbar},
\ee
which is recognized as the non-abelian Fourier expansion corresponding to
$k$ NS5-branes \cite{Pioline:2009qt,Alexandrov:2010ca}, provided we identify $\hbar = 1/k$ and set
the characteristics $(\theta, \phi)$ to zero. This identification of the NS5-brane charge $k$ with
the inverse of a quantization parameter $\hbar$ will be confirmed from
a different perspective in \S\ref{sec_univhyp} below.

Moreover, as explained in \cite{FGquant}, the gluing conditions for
wave functions $\Psi_{\hbar^{-1}, \ell}$ involve a convolution with the quantum dilogarithm.
Since the latter also governs the wall-crossing behavior of motivic DT invariants
$\Omega_{\hbar^{-1}}(\gamma)$ \cite{ks}, this suggests that  NS5-instantons
should be controlled by the same. On the other hand, S-duality implies that
$(p,k)$5-brane instantons are determined by the ordinary DT invariants $\Omega(\gamma)$
suggesting an intriguing relation between ordinary and motivic DT invariants which it would be
very interesting to spell out. Finally, while the analogy with quantum cluster varieties
seems to open the way to the construction of a natural section of $H^1(\cZ,O(2))$, hence a
linear correction to the D-instanton corrected metric on $\cM_H$, it is as yet unclear how
this could be lifted to a full non-linear deformation, consistent
with S-duality and regularity.

\subsection{Universal hypermultiplet and free fermions}
\label{sec_univhyp}

Another interesting relation to quantum integrable systems can be seen  in the special case of type IIA string theory compactified
on a rigid CY threefold with $h^{2,1}(X)=0$. The corresponding HM moduli space is a 4-dimensional QK space, often
known as  `the universal hypermultiplet' (although it is hardly universal).

In the absence of NS5-brane corrections, $\cM_H$ has a continuous isometry corresponding to shifts of the NS-axion $\sigma$.
Four-dimensional QK spaces with an isometry are known to be described by solutions of the Toda equation
\cite{Przanowski:1991ru}
\be
\p_{z} \p_{\bz}  \todaQ +\p_ \rho^2 \, {\rm e}^ \todaQ = 0,
\label{Toda}
\ee
which appears as the lowest equation of the dispersionless (i.e. classical) limit of Toda integrable hierarchy.
In fact, one can show that the twistor framework is equivalent to the Lax formalism for this hierarchy \cite{Guha:1997fz}.
In particular, the Darboux coordinates $\xi$ and $\txi$ coincide with the two Lax operators,
whereas the gluing conditions are identified with the so called string equations.
To understand  the role of the  Darboux coordinate $\alpha$, consider the perturbative
HM moduli space \eqref{pertmetricIIA} with $F=-\tfrac{\I}{4}\, X^2$. In this case the corresponding solution of \eqref{Toda}
takes particularly simple form \cite{Davidse:2005ef,Alexandrov:2006hx}
\be
\todaQ=\log(\rho+c).
\label{solToda}
\ee
In \cite{Alexandrov:2012np} it was shown that the Darboux coordinate $\alpha$ given in \eqref{Darbouxpert}
is related in a simple way to the WKB phase of the quasiclassical Baker-Akhiezer function
$\Psi$ associated with the solution \eqref{solToda}. This, together with the known
wave-function property of the Baker-Akhiezer function, suggests that $\Psi$ might be related to NS5-brane effects,
which typically have an exponential dependence on $\alpha$, see \eqref{5pqZ2}.

Indeed, in \cite{Alexandrov:2009vj} it was suggested that, for compactifications on rigid CY,
NS5-brane instanton corrections to the contact structure on the twistor space
are generated by the following holomorphic function
\be
H^{\rm NS5}_k\sim (\xi - 2\I\txi )^{8\pi c k} \,
e^{- \pi \I k\tilde\alpha -\pi k\(\frac{1}{4}\xi^2+\txi^2\)} ,
\label{NS5fun}
\ee
which is the type IIA counterpart of the type IIB description based on \eqref{5pqZ2}.
One can easily verify that $H^{\rm NS5}_k$ is proportional to the Baker-Akhiezer function $\Psi$
provided one identifies the NS5-brane charge with the inverse quantization parameter
of the integrable hierarchy \cite{Alexandrov:2012np}
\be
\hbar^{-1}=8\pi k.
\ee
This nicely agrees with the relation found below \eqref{Fhbar}, up to normalization.

Furthermore, this 4-dimensional example hints for a possible relation to free fermions. Indeed,
the solution of Toda hierarchy based on \eqref{solToda} is known to describe
non-critical $c=1$ string theory in a non-trivial tachyon background compactified on a circle
of the self-dual radius \cite{Alexandrov:2003ut}.
In turn, this theory is described by Matrix Quantum Mechanics, where integrability is manifest,
and this matrix model is known to reduce to a system of free fermions.
In this description, the Baker-Akhiezer function, shown above to encode NS5-brane effects,
is just the one-fermion wave function. It would be exciting
to use this idea to compute the exact quantum HM moduli space for rigid CY threefolds, and see
whether S-duality or one of its extensions is indeed realized
\cite{Pioline:2009qt,Bao:2009fg,Bao:2010cc,Persson:2011xi}.

\providecommand{\href}[2]{#2}\begingroup\raggedright\endgroup

%\bibliography{../common/combined}

\begin{thebibliography}{100}

\bibitem{Bagger:1983tt}
J.~Bagger and E.~Witten, ``{M}atter couplings in {${\mathcal N}=2$}
  supergravity,'' {\em Nucl. Phys.} {\bf B222} (1983)
1.
%%CITATION = NUPHA,B222,1;%%.

\bibitem{deWit:1984px}
B.~de~Wit, P.~Lauwers, and A.~Van~Proeyen, ``{Lagrangians of N=2 Supergravity -
  Matter Systems},'' {\em Nucl.Phys.} {\bf B255} (1985)
569.
%%CITATION = NUPHA,B255,569;%%.

\bibitem{MR664330}
S.~M. Salamon, ``Quaternionic {K}\"ahler manifolds,'' {\em Invent. Math.} {\bf
  67} (1982), no.~1, 143--171.

\bibitem{Karlhede:1984vr}
A.~Karlhede, U.~Lindstr{\"o}m, and M.~Ro\v{c}ek, ``Selfinteracting tensor
  multiplets in {$\N=2$} superspace,'' {\em Phys. Lett.} {\bf B147} (1984)
297.
%%CITATION = PHLTA,B147,297;%%.

\bibitem{Hitchin:1986ea}
N.~J. Hitchin, A.~Karlhede, U.~Lindstr{\"o}m, and M.~Ro\v{c}ek,
  ``Hyperk{\"a}hler metrics and supersymmetry,'' {\em Commun. Math. Phys.} {\bf
  108} (1987)
535.
%%CITATION = CMPHA,108,535;%%.

\bibitem{MR1001707}
C.~LeBrun, ``Quaternionic-{K}\"ahler manifolds and conformal geometry,'' {\em
  Math. Ann.} {\bf 284} (1989), no.~3, 353--376.

\bibitem{MR1096180}
A.~Swann, ``Hyper-{K}\"ahler and quaternionic {K}\"ahler geometry,'' {\em Math.
  Ann.} {\bf 289} (1991), no.~3, 421--450.

\bibitem{deWit:2001dj}
B.~de~Wit, M.~Ro\v{c}ek, and S.~Vandoren, ``{H}ypermultiplets, hyperk{\"a}hler
  cones and quaternion-{K\"a}hler geometry,'' {\em JHEP} {\bf 02} (2001) 039,
\href{http://www.arXiv.org/abs/hep-th/0101161}{{\tt hep-th/0101161}}.
%%CITATION = HEP-TH 0101161;%%.

\bibitem{Lindstrom:2008gs}
U.~Lindstrom and M.~Rocek, ``{Properties of hyperkahler manifolds and their
  twistor spaces},'' {\em Commun.Math.Phys.} {\bf 293} (2010) 257--278,
  \href{http://www.arXiv.org/abs/0807.1366}{{\tt 0807.1366}}.

\bibitem{Alexandrov:2011va}
S.~Alexandrov, ``{Twistor Approach to String Compactifications: a Review},''
  {\em Phys.Rept.} {\bf 522} (2013) 1--57,
\href{http://www.arXiv.org/abs/1111.2892}{{\tt 1111.2892}}.
%%CITATION = ARXIV:1111.2892;%%.

\bibitem{Louis:2011aa}
J.~Louis and R.~Valandro, ``{Heterotic-Type II Duality in the Hypermultiplet
  Sector},'' {\em JHEP} {\bf 1205} (2012) 016,
\href{http://www.arXiv.org/abs/1112.3566}{{\tt 1112.3566}}.
%%CITATION = ARXIV:1112.3566;%%.

\bibitem{Alexandrov:2012pr}
S.~Alexandrov and B.~Pioline, ``{Heterotic-type II duality in twistor space},''
  {\em JHEP} {\bf 1303} (2013) 085,
\href{http://www.arXiv.org/abs/1210.3037}{{\tt 1210.3037}}.
%%CITATION = ARXIV:1210.3037;%%.

\bibitem{Alexandrov:2014jua}
S.~Alexandrov, J.~Louis, B.~Pioline, and R.~Valandro, ``{$\mathcal N=2$
  Heterotic-Type II duality and bundle moduli},'' {\em JHEP} {\bf 1408} (2014)
  092,
\href{http://www.arXiv.org/abs/1405.4792}{{\tt 1405.4792}}.
%%CITATION = ARXIV:1405.4792;%%.

\bibitem{Gunther:1998sc}
H.~G{\"u}nther, C.~Herrmann, and J.~Louis, ``{Quantum corrections in the
  hypermultiplet moduli space},'' {\em Fortsch. Phys.} {\bf 48} (2000)
  119--123,
\href{http://www.arXiv.org/abs/hep-th/9901137}{{\tt hep-th/9901137}}.
%%CITATION = HEP-TH/9901137;%%.

\bibitem{Robles-Llana:2006ez}
D.~Robles-Llana, F.~Saueressig, and S.~Vandoren, ``String loop corrected
  hypermultiplet moduli spaces,'' {\em JHEP} {\bf 03} (2006) 081,
\href{http://www.arXiv.org/abs/hep-th/0602164}{{\tt hep-th/0602164}}.
%%CITATION = HEP-TH 0602164;%%.

\bibitem{Alexandrov:2008gh}
S.~Alexandrov, B.~Pioline, F.~Saueressig, and S.~Vandoren, ``{D-instantons and
  twistors},'' {\em JHEP} {\bf 03} (2009) 044,
\href{http://www.arXiv.org/abs/0812.4219}{{\tt 0812.4219}}.
%%CITATION = 0812.4219;%%.

\bibitem{Candelas:1990rm}
P.~Candelas, X.~C. de~la Ossa, P.~S. Green, and L.~Parkes, ``{A pair of
  Calabi-Yau manifolds as an exactly soluble superconformal theory},'' {\em
  Nucl. Phys.} {\bf B359} (1991)
21--74.
%%CITATION = NUPHA,B359,21;%%.

\bibitem{Alexandrov:2010np}
S.~Alexandrov, D.~Persson, and B.~Pioline, ``{On the topology of the
  hypermultiplet moduli space in type II/CY string vacua},'' {\em Phys.Rev.}
  {\bf D83} (2011) 026001, \href{http://www.arXiv.org/abs/1009.3026}{{\tt
  1009.3026}}.

\bibitem{Alexandrov:2010ca}
S.~Alexandrov, D.~Persson, and B.~Pioline, ``{Fivebrane instantons, topological
  wave functions and hypermultiplet moduli spaces},'' {\em JHEP} {\bf 1103}
  (2011) 111, \href{http://www.arXiv.org/abs/1010.5792}{{\tt 1010.5792}}.

\bibitem{Cecotti:1989qn}
S.~Cecotti, S.~Ferrara, and L.~Girardello, ``Geometry of type {I}{I}
  superstrings and the moduli of superconformal field theories,'' {\em Int. J.
  Mod. Phys.} {\bf A4} (1989)
2475.
%%CITATION = IMPAE,A4,2475;%%.

\bibitem{Ferrara:1989ik}
S.~Ferrara and S.~Sabharwal, ``{Q}uaternionic manifolds for type {II}
  superstring vacua of {C}alabi-{Y}au spaces,'' {\em Nucl. Phys.} {\bf B332}
  (1990)
317.
%%CITATION = NUPHA,B332,317;%%.

\bibitem{Strominger:1990pd}
A.~Strominger, ``{SPECIAL GEOMETRY},'' {\em Commun.Math.Phys.} {\bf 133} (1990)
163--180.
%%CITATION = CMPHA,133,163;%%.

\bibitem{Antoniadis:2003sw}
I.~Antoniadis, R.~Minasian, S.~Theisen, and P.~Vanhove, ``String loop
  corrections to the universal hypermultiplet,'' {\em Class. Quant. Grav.} {\bf
  20} (2003) 5079--5102,
\href{http://www.arXiv.org/abs/hep-th/0307268}{{\tt hep-th/0307268}}.
%%CITATION = HEP-TH/0307268;%%.

\bibitem{Alexandrov:2007ec}
S.~Alexandrov, ``{Quantum covariant c-map},'' {\em JHEP} {\bf 05} (2007) 094,
\href{http://www.arXiv.org/abs/hep-th/0702203}{{\tt hep-th/0702203}}.
%%CITATION = HEP-TH/0702203;%%.

\bibitem{Alexandrov:2014sya}
S.~Alexandrov and S.~Banerjee, ``{Hypermultiplet metric and D-instantons},''
\href{http://www.arXiv.org/abs/1412.8182}{{\tt 1412.8182}}.
%%CITATION = ARXIV:1412.8182;%%.

\bibitem{Neitzke:2007ke}
A.~Neitzke, B.~Pioline, and S.~Vandoren, ``{Twistors and Black Holes},'' {\em
  JHEP} {\bf 04} (2007) 038,
\href{http://www.arXiv.org/abs/hep-th/0701214}{{\tt hep-th/0701214}}.
%%CITATION = HEP-TH/0701214;%%.

\bibitem{Alexandrov:2008nk}
S.~Alexandrov, B.~Pioline, F.~Saueressig, and S.~Vandoren, ``{Linear
  perturbations of quaternionic metrics},'' {\em Commun. Math. Phys.} {\bf 296}
  (2010) 353--403,
\href{http://www.arXiv.org/abs/0810.1675}{{\tt 0810.1675}}.
%%CITATION = 0810.1675;%%.

\bibitem{Rocek:2005ij}
M.~Ro\v{c}ek, C.~Vafa, and S.~Vandoren, ``Hypermultiplets and topological
  strings,'' {\em JHEP} {\bf 02} (2006) 062,
\href{http://www.arXiv.org/abs/hep-th/0512206}{{\tt hep-th/0512206}}.
%%CITATION = HEP-TH 0512206;%%.

\bibitem{MR872143}
K.~Galicki, ``A generalization of the momentum mapping construction for
  quaternionic {K}\"ahler manifolds,'' {\em Comm. Math. Phys.} {\bf 108}
  (1987), no.~1, 117--138.

\bibitem{lebrun1994srp}
C.~LeBrun and S.~Salamon, ``{Strong rigidity of positive quaternion-K{\"a}hler
  manifolds},'' {\em Inventiones Mathematicae} {\bf 118} (1994), no.~1,
  109--132.

\bibitem{Louis:2002ny}
J.~Louis and A.~Micu, ``{Type II theories compactified on Calabi-Yau threefolds
  in the presence of background fluxes},'' {\em Nucl. Phys.} {\bf B635} (2002)
  395--431,
\href{http://www.arXiv.org/abs/hep-th/0202168}{{\tt hep-th/0202168}}.
%%CITATION = HEP-TH/0202168;%%.

\bibitem{Bohm:1999uk}
R.~B{\"o}hm, H.~G{\"u}nther, C.~Herrmann, and J.~Louis, ``{Compactification of
  type IIB string theory on Calabi-Yau threefolds},'' {\em Nucl. Phys.} {\bf
  B569} (2000) 229--246,
\href{http://www.arXiv.org/abs/hep-th/9908007}{{\tt hep-th/9908007}}.
%%CITATION = HEP-TH/9908007;%%.

\bibitem{RoblesLlana:2006is}
D.~Robles-Llana, M.~Ro\v{c}ek, F.~Saueressig, U.~Theis, and S.~Vandoren,
  ``{Nonperturbative corrections to 4D string theory effective actions from
  SL(2,Z) duality and supersymmetry},'' {\em Phys. Rev. Lett.} {\bf 98} (2007)
  211602,
\href{http://www.arXiv.org/abs/hep-th/0612027}{{\tt hep-th/0612027}}.
%%CITATION = HEP-TH/0612027;%%.

\bibitem{Alexandrov:2009qq}
S.~Alexandrov and F.~Saueressig, ``{Quantum mirror symmetry and twistors},''
  {\em JHEP} {\bf 09} (2009) 108,
\href{http://www.arXiv.org/abs/0906.3743}{{\tt 0906.3743}}.
%%CITATION = 0906.3743;%%.

\bibitem{Alexandrov:2012bu}
S.~Alexandrov and B.~Pioline, ``{S-duality in Twistor Space},'' {\em JHEP} {\bf
  1208} (2012) 112,
\href{http://www.arXiv.org/abs/1206.1341}{{\tt 1206.1341}}.
%%CITATION = ARXIV:1206.1341;%%.

\bibitem{Green:1997tv}
M.~B. Green and M.~Gutperle, ``{Effects of D-instantons},'' {\em Nucl. Phys.}
  {\bf B498} (1997) 195--227,
\href{http://www.arXiv.org/abs/hep-th/9701093}{{\tt hep-th/9701093}}.
%%CITATION = HEP-TH/9701093;%%.

\bibitem{Alexandrov:2012au}
S.~Alexandrov, J.~Manschot, and B.~Pioline, ``{D3-instantons, Mock Theta Series
  and Twistors},'' {\em JHEP} {\bf 1304} (2013) 002, 
\href{http://www.arXiv.org/abs/1207.1109}{{\tt 1207.1109}}.
%%CITATION = ARXIV:1207.1109;%%.

\bibitem{Sharpe:1999qz}
E.~R. Sharpe, ``{D-branes, derived categories, and Grothendieck groups},'' {\em
  Nucl.Phys.} {\bf B561} (1999) 433--450,
\href{http://www.arXiv.org/abs/hep-th/9902116}{{\tt hep-th/9902116}}.
%%CITATION = HEP-TH/9902116;%%.

\bibitem{Douglas:2000ah}
M.~R. Douglas, B.~Fiol, and C.~Romelsberger, ``{Stability and BPS branes},''
  {\em JHEP} {\bf 0509} (2005) 006,
\href{http://www.arXiv.org/abs/hep-th/0002037}{{\tt hep-th/0002037}}.
%%CITATION = HEP-TH/0002037;%%.

\bibitem{MR1403918}
M.~Kontsevich, ``Homological algebra of mirror symmetry,'' in {\em Proceedings
  of the {I}nternational {C}ongress of {M}athematicians, {V}ol.\ 1, 2
  ({Z}\"urich, 1994)}, pp.~120--139.
\newblock Birkh\"auser, Basel, 1995.
\newblock \href{http://www.arXiv.org/abs/alg-geom/9411018}{{\tt
  alg-geom/9411018}}.

\bibitem{Ferrara:1995yx}
S.~Ferrara, J.~A. Harvey, A.~Strominger, and C.~Vafa, ``Second quantized mirror
  symmetry,'' {\em Phys. Lett.} {\bf B361} (1995) 59--65,
\href{http://arXiv.org/abs/hep-th/9505162}{{\tt hep-th/9505162}}.
%%CITATION = HEP-TH 9505162;%%.

\bibitem{DouglasTopMirrorMap}
M.~R. Douglas, ``{Dirichlet branes, homological mirror symmetry, and
  stability},'' {\em ICM proceedings} (2002)
  \href{http://www.arXiv.org/abs/math/0207021}{{\tt math/0207021}}.

\bibitem{MR2373143}
T.~Bridgeland, ``Stability conditions on triangulated categories,'' {\em Ann.
  of Math. (2)} {\bf 166} (2007), no.~2, 317--345.

\bibitem{MR666108}
R.~Harvey and H.~B. Lawson, Jr., ``Calibrated geometries,'' {\em Acta Math.}
  {\bf 148} (1982) 47--157.

\bibitem{Andriyash:2010qv}
E.~Andriyash, F.~Denef, D.~L. Jafferis, and G.~W. Moore, ``{Wall-crossing from
  supersymmetric galaxies},'' {\em JHEP} {\bf 1201} (2012) 115,
\href{http://www.arXiv.org/abs/1008.0030}{{\tt 1008.0030}}.
%%CITATION = ARXIV:1008.0030;%%.

\bibitem{Pioline:2011gf}
B.~Pioline, ``{Four ways across the wall},'' {\em J.Phys.Conf.Ser.} {\bf 346}
  (2012) 012017,
\href{http://www.arXiv.org/abs/1103.0261}{{\tt 1103.0261}}.
%%CITATION = ARXIV:1103.0261;%%.

\bibitem{ks}
M.~Kontsevich and Y.~Soibelman, ``{Stability structures, motivic
  Donaldson-Thomas invariants and cluster transformations},''
  \href{http://www.arXiv.org/abs/0811.2435}{{\tt 0811.2435}}.

\bibitem{Gaiotto:2008cd}
D.~Gaiotto, G.~W. Moore, and A.~Neitzke, ``{Four-dimensional wall-crossing via
  three-dimensional field theory},'' {\em Commun.Math.Phys.} {\bf 299} (2010)
  163--224, \href{http://www.arXiv.org/abs/0807.4723}{{\tt 0807.4723}}.

\bibitem{Joyce:2008pc}
D.~Joyce and Y.~Song, ``{A theory of generalized Donaldson-Thomas
  invariants},''
\href{http://www.arXiv.org/abs/0810.5645}{{\tt 0810.5645}}.
%%CITATION = 0810.5645;%%.

\bibitem{Manschot:2010qz}
J.~Manschot, B.~Pioline, and A.~Sen, ``{Wall Crossing from Boltzmann Black Hole
  Halos},'' {\em JHEP} {\bf 1107} (2011) 059,
\href{http://www.arXiv.org/abs/1011.1258}{{\tt 1011.1258}}.
%%CITATION = ARXIV:1011.1258;%%.

\bibitem{Alexandrov:2009zh}
S.~Alexandrov, ``{D-instantons and twistors: some exact results},'' {\em J.
  Phys.} {\bf A42} (2009) 335402,
\href{http://www.arXiv.org/abs/0902.2761}{{\tt 0902.2761}}.
%%CITATION = 0902.2761;%%.

\bibitem{Alexandrov:2011ac}
S.~Alexandrov, D.~Persson, and B.~Pioline, ``{Wall-crossing, Rogers
  dilogarithm, and the QK/HK correspondence},'' {\em JHEP} {\bf 1112} (2011)
  027,
\href{http://www.arXiv.org/abs/1110.0466}{{\tt 1110.0466}}.
%%CITATION = ARXIV:1110.0466;%%.

\bibitem{Stoppa:2011}
J.~Stoppa, ``{Joyce-Song wall-crossing as an asymptotic expansion},''
{\em Kyoto J. Math.} {\bf 54} (2014) 103-156,
\href{http://www.arXiv.org/abs/1112.2174}{{\tt 1112.2174}}.
%%CITATION = 1112.2174;%%.

\bibitem{Alexandrov:2014wca} 
  S.~Alexandrov, G.~W.~Moore, A.~Neitzke and B.~Pioline,
  ``An $R^3$ index for four-dimensional $N=2$ field theories,''
\href{http://www.arXiv.org/abs/1112.2174}{{\tt 1406.2360}}. 

\bibitem{MR967469}
M.~M. Capria and S.~M. Salamon, ``Yang-{M}ills fields on quaternionic spaces,''
  {\em Nonlinearity} {\bf 1} (1988), no.~4, 517--530.

\bibitem{MR1139657}
T.~Gocho and H.~Nakajima, ``Einstein-{H}ermitian connections on
  hyper-{K}\"ahler quotients,'' {\em J. Math. Soc. Japan} {\bf 44} (1992),
  no.~1, 43--51.

\bibitem{MR1486984}
M.~Verbitsky, ``Hyperholomorphic bundles over a hyper-{K}\"ahler manifold,''
  {\em J. Algebraic Geom.} {\bf 5} (1996), no.~4, 633--669.

\bibitem{Haydys}
A.~Haydys, ``{Hyper-K\"ahler and quaternionic K\"ahler manifolds with
  $S^{1}$-symmetries},'' {\em J. Geom. Phys.} {\bf 58} (2008), no.~3, 293--306.

\bibitem{Hitchin:2012}
N.~Hitchin, ``{On the hyperkaehler/quaternion Kaehler
  correspondence},'' {\em Comm. Math. Phys.} {\bf 324} (2013) 77-106,
\href{http://www.arXiv.org/abs/1210.0424}{{\tt 1210.0424}}.
%%CITATION = HEP-TH/0011220;%%.

\bibitem{Alekseevsky:2012fu}
D.~V. Alekseevsky, V.~Cortes, and T.~Mohaupt, ``{Conification of K\"ahler and
  hyper-K\"ahler manifolds},''
\href{http://www.arXiv.org/abs/1205.2964}{{\tt 1205.2964}}.
%%CITATION = ARXIV:1205.2964;%%.

\bibitem{ward-wells}
R.~S. Ward and R.~O. Wells, {\em {T}wistor geometry and field theory}.
\newblock Cambridge Monographs on Mathematical Physics. Cambridge University
  Press, Cambridge, 1990.

\bibitem{Neitzke:2011za}
A.~Neitzke, ``{On a hyperholomorphic line bundle over the Coulomb branch},''
\href{http://www.arXiv.org/abs/1110.1619}{{\tt 1110.1619}}.
%%CITATION = 1110.1619;%%.

\bibitem{Gliozzi:1994cs}
F.~Gliozzi and R.~Tateo, ``{ADE functional dilogarithm identities and
  integrable models},'' {\em Phys.Lett.} {\bf B348} (1995) 84--88,
\href{http://www.arXiv.org/abs/hep-th/9411203}{{\tt hep-th/9411203}}.
%%CITATION = HEP-TH/9411203;%%.

\bibitem{MR2804544}
T.~Nakanishi, ``Dilogarithm identities for conformal field theories and cluster
  algebras: simply laced case,'' {\em Nagoya Math. J.} {\bf 202} (2011) 23--43.

\bibitem{Pioline:2009ia}
B.~Pioline and S.~Vandoren, ``{Large D-instanton effects in string theory},''
  {\em JHEP} {\bf 07} (2009) 008,
\href{http://www.arXiv.org/abs/0904.2303}{{\tt 0904.2303}}.
%%CITATION = 0904.2303;%%.

\bibitem{Alexandrov:2013mha}
S.~Alexandrov and S.~Banerjee, ``{Modularity, quaternion-KŠhler spaces, and
  mirror symmetry},'' {\em J.Math.Phys.} {\bf 54} (2013) 102301,
\href{http://www.arXiv.org/abs/1306.1837}{{\tt 1306.1837}}.
%%CITATION = ARXIV:1306.1837;%%.

\bibitem{Alexandrov:2014mfa}
S.~Alexandrov and S.~Banerjee, ``{Fivebrane instantons in Calabi-Yau
  compactifications},'' {\em Phys.Rev.} {\bf D90} (2014) 041902,
\href{http://www.arXiv.org/abs/1403.1265}{{\tt 1403.1265}}.
%%CITATION = ARXIV:1403.1265;%%.

\bibitem{Maldacena:1997de}
J.~M. Maldacena, A.~Strominger, and E.~Witten, ``{B}lack hole entropy in
  {M}-theory,'' {\em JHEP} {\bf 12} (1997) 002,
\href{http://www.arXiv.org/abs/hep-th/9711053}{{\tt hep-th/9711053}}.
%%CITATION = HEP-TH 9711053;%%.

\bibitem{Gaiotto:2006ns}
D.~Gaiotto, A.~Strominger, and X.~Yin, ``From {A}d{S}(3)/{CFT}(2) to black
  holes / topological strings,''
\href{http://www.arXiv.org/abs/hep-th/0602046}{{\tt hep-th/0602046}}.
%%CITATION = HEP-TH 0602046;%%.

\bibitem{Gaiotto:2006wm}
D.~Gaiotto, A.~Strominger, and X.~Yin, ``{The M5-brane elliptic genus:
  Modularity and BPS states},'' {\em JHEP} {\bf 08} (2007) 070,
\href{http://www.arXiv.org/abs/hep-th/0607010}{{\tt hep-th/0607010}}.
%%CITATION = HEP-TH/0607010;%%.

\bibitem{deBoer:2006vg}
J.~de~Boer, M.~C.~N. Cheng, R.~Dijkgraaf, J.~Manschot, and E.~Verlinde, ``{A
  farey tail for attractor black holes},'' {\em JHEP} {\bf 11} (2006) 024,
\href{http://www.arXiv.org/abs/hep-th/0608059}{{\tt hep-th/0608059}}.
%%CITATION = HEP-TH/0608059;%%.

\bibitem{Denef:2007vg}
F.~Denef and G.~W. Moore, ``{Split states, entropy enigmas, holes and halos},''
  {\em JHEP} {\bf 1111} (2011) 129,
\href{http://www.arXiv.org/abs/hep-th/0702146}{{\tt hep-th/0702146}}.
%%CITATION = HEP-TH/0702146;%%.

\bibitem{Dabholkar:2005dt}
A.~Dabholkar, F.~Denef, G.~W. Moore, and B.~Pioline, ``Precision counting of
  small black holes,'' {\em JHEP} {\bf 10} (2005) 096,
\href{http://www.arXiv.org/abs/hep-th/0507014}{{\tt hep-th/0507014}}.
%%CITATION = HEP-TH/0507014;%%.

\bibitem{Manschot:2009ia}
J.~Manschot, ``{Stability and duality in N=2 supergravity},'' {\em
  Commun.Math.Phys.} {\bf 299} (2010) 651--676,
  \href{http://www.arXiv.org/abs/0906.1767}{{\tt 0906.1767}}.

\bibitem{deBoer:2008fk}
J.~de~Boer, F.~Denef, S.~El-Showk, I.~Messamah, and D.~Van~den Bleeken,
  ``{Black hole bound states in $AdS_3 \times S^2$},'' {\em JHEP} {\bf 0811}
  (2008) 050,
\href{http://www.arXiv.org/abs/0802.2257}{{\tt 0802.2257}}.
%%CITATION = ARXIV:0802.2257;%%.

\bibitem{Manschot:2010xp}
J.~Manschot, ``{Wall-crossing of D4-branes using flow trees},'' {\em
  Adv.Theor.Math.Phys.} {\bf 15} (2011) 1--42,
\href{http://www.arXiv.org/abs/1003.1570}{{\tt 1003.1570}}.
%%CITATION = ARXIV:1003.1570;%%.

\bibitem{Manschot:2011dj}
J.~Manschot, ``{BPS invariants of N=4 gauge theory on a surface},'' {\em
  Commun. Num. Theor. Phys.} {\bf 06} (2012) 497--516,
\href{http://www.arXiv.org/abs/1103.0012}{{\tt 1103.0012}}.
%%CITATION = ARXIV:1103.0012;%%.

\bibitem{Troost:2010ud}
J.~Troost, ``{The non-compact elliptic genus: mock or modular},'' {\em JHEP}
  {\bf 1006} (2010) 104,
\href{http://www.arXiv.org/abs/1004.3649}{{\tt 1004.3649}}.
%%CITATION = ARXIV:1004.3649;%%.

\bibitem{Vafa:1994tf}
C.~Vafa and E.~Witten, ``{A Strong coupling test of S duality},'' {\em
  Nucl.Phys.} {\bf B431} (1994) 3--77,
\href{http://www.arXiv.org/abs/hep-th/9408074}{{\tt hep-th/9408074}}.
%%CITATION = HEP-TH/9408074;%%.

\bibitem{Zwegers-thesis}
S.~Zwegers, ``Mock theta functions.'' PhD dissertation, 2002, Utrecht.

\bibitem{Becker:1995kb}
K.~Becker, M.~Becker, and A.~Strominger, ``Five-branes, membranes and
  nonperturbative string theory,'' {\em Nucl. Phys.} {\bf B456} (1995)
  130--152,
\href{http://www.arXiv.org/abs/hep-th/9507158}{{\tt hep-th/9507158}}.
%%CITATION = NUPHA,B456,130;%%.

\bibitem{Witten:1996hc}
E.~Witten, ``{Five-brane effective action in M-theory},'' {\em J. Geom. Phys.}
  {\bf 22} (1997) 103--133,
\href{http://www.arXiv.org/abs/hep-th/9610234}{{\tt hep-th/9610234}}.
%%CITATION = HEP-TH/9610234;%%.

\bibitem{Henningson:1999dm}
M.~Henningson, B.~E.~W. Nilsson, and P.~Salomonson, ``{Holomorphic
  factorization of correlation functions in (4k+2)-dimensional (2k)-form gauge
  theory},'' {\em JHEP} {\bf 09} (1999) 008,
\href{http://www.arXiv.org/abs/hep-th/9908107}{{\tt hep-th/9908107}}.
%%CITATION = HEP-TH/9908107;%%.

\bibitem{Dijkgraaf:2002ac}
R.~Dijkgraaf, E.~P. Verlinde, and M.~Vonk, ``{O}n the partition sum of the {NS}
  five-brane,''
\href{http://www.arXiv.org/abs/hep-th/0205281}{{\tt hep-th/0205281}}.
%%CITATION = HEP-TH 0205281;%%.

\bibitem{Moore:2004jv}
G.~W. Moore, ``{Anomalies, Gauss laws, and Page charges in M-theory},'' {\em
  Comptes Rendus Physique} {\bf 6} (2005) 251--259,
\href{http://www.arXiv.org/abs/hep-th/0409158}{{\tt hep-th/0409158}}.
%%CITATION = HEP-TH/0409158;%%.

\bibitem{Belov:2006jd}
D.~Belov and G.~W. Moore, ``{Holographic action for the self-dual field},''
\href{http://www.arXiv.org/abs/hep-th/0605038}{{\tt hep-th/0605038}}.
%%CITATION = HEP-TH/0605038;%%.

\bibitem{Duff:1995wd}
M.~Duff, J.~T. Liu, and R.~Minasian, ``{Eleven-dimensional origin of
  string-string duality: A One loop test},'' {\em Nucl.Phys.} {\bf B452} (1995)
  261--282,
\href{http://www.arXiv.org/abs/hep-th/9506126}{{\tt hep-th/9506126}}.
%%CITATION = HEP-TH/9506126;%%.

\bibitem{Witten:1985xe}
E.~Witten, ``{Global gravitational anomalies},'' {\em Commun.Math.Phys.} {\bf
  100} (1985)
197.
%%CITATION = CMPHA,100,197;%%.

\bibitem{Monnier:2010ww}
S.~Monnier, ``{Geometric quantization and the metric dependence of the
  self-dual field theory},'' {\em Commun.Math.Phys.} {\bf 314} (2012) 305--328,
\href{http://www.arXiv.org/abs/1011.5890}{{\tt 1011.5890}}.
%%CITATION = ARXIV:1011.5890;%%.

\bibitem{1167.32016}
H.~Fang, Z.~Lu, and K.-I. Yoshikawa, ``{Analytic torsion for Calabi-Yau
  threefolds.},'' {\em J. Differ. Geom.} {\bf 80} (2008), no.~2, 175--259.

\bibitem{2012arXiv1212.3790H}
Z.~{Hua}, ``{Spin structure on moduli space of sheaves on Calabi-Yau
  threefold},'' {\em ArXiv e-prints} (Dec., 2012)
  \href{http://www.arXiv.org/abs/1212.3790}{{\tt 1212.3790}}.

\bibitem{Looyestijn:2008pg}
H.~Looyestijn and S.~Vandoren, ``{On NS5-brane instantons and volume
  stabilization},'' {\em JHEP} {\bf 0804} (2008) 024,
  \href{http://www.arXiv.org/abs/0801.3949}{{\tt 0801.3949}}.

\bibitem{Alexandrov:2014rca}
S.~Alexandrov and S.~Banerjee, ``{Dualities and fivebrane instantons},'' {\em
  JHEP} {\bf 1411} (2014) 040,
\href{http://www.arXiv.org/abs/1405.0291}{{\tt 1405.0291}}.
%%CITATION = ARXIV:1405.0291;%%.

\bibitem{gw-dt}
D.~Maulik, N.~Nekrasov, A.~Okounkov, and R.~Pandharipande, ``Gromov-{W}itten
  theory and {D}onaldson-{T}homas theory. {I},'' {\em Compos. Math.} {\bf 142}
  (2006), no.~5, 1263--1285.

\bibitem{Nekrasov:2004js}
N.~Nekrasov, H.~Ooguri, and C.~Vafa, ``{S duality and topological strings},''
  {\em JHEP} {\bf 0410} (2004) 009,
  \href{http://www.arXiv.org/abs/hep-th/0403167}{{\tt hep-th/0403167}}.

\bibitem{Kapustin:2004jm}
A.~Kapustin, ``{G}auge theory, topological strings, and {S}-duality,'' {\em
  JHEP} {\bf 09} (2004) 034,
\href{http://www.arXiv.org/abs/hep-th/0404041}{{\tt hep-th/0404041}}.
%%CITATION = HEP-TH 0404041;%%.

\bibitem{2013arXiv1303.3253K}
M.~{Kontsevich} and Y.~{Soibelman}, ``{Wall-crossing structures in
  Donaldson-Thomas invariants, integrable systems and Mirror Symmetry},'' {\em
  ArXiv e-prints} (Mar., 2013) \href{http://www.arXiv.org/abs/1303.3253}{{\tt
  1303.3253}}.

\bibitem{Alexandrov:2010pp}
S.~Alexandrov and P.~Roche, ``{TBA for non-perturbative moduli spaces},'' {\em
  JHEP} {\bf 1006} (2010) 066, \href{http://www.arXiv.org/abs/1003.3964}{{\tt
  1003.3964}}.

\bibitem{Zamolodchikov:1989cf}
A.~B. Zamolodchikov, ``{Thermodynamic Bethe Ansatz in Relativistic Models.
  Scaling Three State Potts and Lee-Yang Models},'' {\em Nucl. Phys.} {\bf
  B342} (1990)
695--720.
%%CITATION = NUPHA,B342,695;%%.

\bibitem{Ravanini:1992fi}
F.~Ravanini, R.~Tateo, and A.~Valleriani, ``{Dynkin TBAs},'' {\em Int. J. Mod.
  Phys.} {\bf A8} (1993) 1707--1728,
\href{http://www.arXiv.org/abs/hep-th/9207040}{{\tt hep-th/9207040}}.
%%CITATION = HEP-TH/9207040;%%.

\bibitem{Gaiotto:2009hg}
D.~Gaiotto, G.~W. Moore, and A.~Neitzke, ``{Wall-crossing, Hitchin Systems, and
  the WKB Approximation},''
\href{http://www.arXiv.org/abs/0907.3987}{{\tt 0907.3987}}.
%%CITATION = 0907.3987;%%.

\bibitem{2001math4151F}
S.~{Fomin} and A.~{Zelevinsky}, ``{Cluster algebras I: Foundations},'' {\em
  ArXiv Mathematics e-prints} (Apr., 2001)
  \href{http://www.arXiv.org/abs/arXiv:math/0104151}{{\tt arXiv:math/0104151}}.

\bibitem{2008InMat.175..223F}
V.~V. {Fock} and A.~B. {Goncharov}, ``{The quantum dilogarithm and
  representations of quantum cluster varieties},'' {\em Inventiones
  Mathematicae} {\bf 175} (Sept., 2008) 223--286,
  \href{http://www.arXiv.org/abs/arXiv:math/0702397}{{\tt arXiv:math/0702397}}.

\bibitem{FGquant}
V.~Fock and A.~Goncharov, ``unpublished.''
\newblock talk by V. Fock at the CQGM meeting in Aarhus, summer 2012.

\bibitem{Pioline:2009qt}
B.~Pioline and D.~Persson, ``{The automorphic NS5-brane},'' {\em Commun. Num.
  Th. Phys.} {\bf 3} (2009), no.~4, 697--754,
\href{http://www.arXiv.org/abs/0902.3274}{{\tt 0902.3274}}.
%%CITATION = 0902.3274;%%.

\bibitem{Przanowski:1991ru}
M.~Przanowski, ``{Killing vector fields in selfdual, Euclidean Einstein spaces
  with Lambda not equal 0},'' {\em J. Math. Phys.} {\bf 32} (1991)
1004--1010.
%%CITATION = JMAPA,32,1004;%%.

\bibitem{Guha:1997fz}
P.~Guha and K.~Takasaki, ``{Dispersionless hierarchies, Hamilton-Jacobi theory
  and twistor correspondences},'' {\em J.Geom.Phys.} (1998) 25,326,
  \href{http://www.arXiv.org/abs/solv-int/9705013}{{\tt solv-int/9705013}}.

\bibitem{Davidse:2005ef}
M.~Davidse, F.~Saueressig, U.~Theis, and S.~Vandoren, ``{Membrane instantons
  and de Sitter vacua},'' {\em JHEP} {\bf 09} (2005) 065,
\href{http://www.arXiv.org/abs/hep-th/0506097}{{\tt hep-th/0506097}}.
%%CITATION = HEP-TH/0506097;%%.

\bibitem{Alexandrov:2006hx}
S.~Alexandrov, F.~Saueressig, and S.~Vandoren, ``{Membrane and fivebrane
  instantons from quaternionic geometry},'' {\em JHEP} {\bf 09} (2006) 040,
\href{http://www.arXiv.org/abs/hep-th/0606259}{{\tt hep-th/0606259}}.
%%CITATION = HEP-TH/0606259;%%.

\bibitem{Alexandrov:2012np}
S.~Alexandrov, ``{c-map as c=1 string},'' {\em Nucl.Phys.} {\bf B863} (2012)
  329--346,
\href{http://www.arXiv.org/abs/1201.4392}{{\tt 1201.4392}}.
%%CITATION = ARXIV:1201.4392;%%.

\bibitem{Alexandrov:2009vj}
S.~Alexandrov, B.~Pioline, and S.~Vandoren, ``{Self-dual Einstein Spaces,
  Heavenly Metrics and Twistors},'' {\em J.Math.Phys.} {\bf 51} (2010) 073510,
  \href{http://www.arXiv.org/abs/0912.3406}{{\tt 0912.3406}}.

\bibitem{Alexandrov:2003ut}
S.~Alexandrov, ``{Matrix quantum mechanics and two-dimensional string theory in
  non-trivial backgrounds},''
\href{http://www.arXiv.org/abs/hep-th/0311273}{{\tt hep-th/0311273}}.
%%CITATION = HEP-TH/0311273;%%.

\bibitem{Bao:2009fg}
L.~Bao, A.~Kleinschmidt, B.~E.~W. Nilsson, D.~Persson, and B.~Pioline,
  ``{Instanton Corrections to the Universal Hypermultiplet and Automorphic
  Forms on SU(2,1)},'' {\em Commun. Num. Theor. Phys.} {\bf 4} (2010) 187--266,
\href{http://www.arXiv.org/abs/0909.4299}{{\tt 0909.4299}}.
%%CITATION = 0909.4299;%%.

\bibitem{Bao:2010cc}
L.~Bao, A.~Kleinschmidt, B.~E.~W. Nilsson, D.~Persson, and B.~Pioline, ``{Rigid
  Calabi-Yau threefolds, Picard Eisenstein series and instantons},''
{\em J. Phys. Conf.  Ser.} {\bf 462} (2013) 012026,
\href{http://www.arXiv.org/abs/1005.4848}{{\tt 1005.4848}}.
%%CITATION = 1005.4848;%%.

\bibitem{Persson:2011xi}
D.~Persson, ``{Automorphic Instanton Partition Functions on Calabi-Yau
  Threefolds},'' {\em J.Phys.Conf.Ser.} {\bf 346} (2012) 012016,
\href{http://www.arXiv.org/abs/1103.1014}{{\tt 1103.1014}}.
%%CITATION = ARXIV:1103.1014;%%.

\end{thebibliography}
%\bibliographystyle{../common/utphys}

\end{document}